\documentclass[showpacs,preprintnumbers,amsmath,aps,nofootinbib,amssymb,superscriptaddress]{revtex4}
\usepackage{graphicx}
\usepackage{dcolumn}
\usepackage[english]{babel}
\usepackage{bm}
\usepackage[font=scriptsize]{caption}
\usepackage{ragged2e}
\usepackage[font=small,labelfont=bf,justification=raggedright]{caption}
\usepackage[utf8]{inputenc}
\usepackage[tikz]{bclogo}
\usepackage[framemethod=tikz]{mdframed}
\usepackage{amsmath}

\usepackage{mathtools}
\usepackage{xcolor}
\usepackage{lscape}
\usepackage{rotating}
\colorlet{BLUE}{blue} \colorlet{RED}{red}
\newsavebox{\measurebox}

\begin{document}

\title[]{Quintom fields from chiral K-essence cosmology}
\author{J. Socorro}
\email{socorro@fisica.ugto.mx}
 \affiliation{Departamento de
F\'{\i}sica, DCeI, Universidad de Guanajuato-Campus Le\'on, C.P.
37150, Le\'on, Guanajuato, M\'exico}

\author{S. P\'erez-Pay\'an}
\email{saperezp@ipn.mx} \affiliation{Unidad Profesional
Interdisciplinaria de Ingenier\'ia,
Campus Guana\-jua\-to del Instituto Polit\'ecnico Nacional.\\
Av. Mineral de Valenciana \#200, Col. Fraccionamiento Industrial
Puerto Interior, C.P. 36275, Silao de la Victoria, Guana\-jua\.to,
M\'exico.}

\author{Rafael Hern\'andez-Jim\'enez}
\email{rafaelhernandezjmz@gmail.com}
 \affiliation{Departamento de
F\'isica, Centro Universitario de Ciencias
Exactas e Ingenier\'ia, Universidad de Guadalajara.\\
Av. Revoluci\'on 1500, Colonia Ol\'impica C.P. 44430, Guadalajara,
Jalisco, M\'exico.}

\author{Abraham Espinoza-Garc\'ia}
\email{aespinoza@ipn.mx} \affiliation{Unidad Profesional
Interdisciplinaria de Ingenier\'ia,
Campus Guana\-jua\-to del Instituto Polit\'ecnico Nacional.\\
Av. Mineral de Valenciana \#200, Col. Fraccionamiento Industrial
Puerto Interior, C.P. 36275, Silao de la Victoria, Guana\-jua\.to,
M\'exico.}

\author{Luis Rey D\'iaz-Barr\'on}
\email{lrdiaz@ipn.mx} \affiliation{Unidad Profesional
Interdisciplinaria de Ingenier\'ia,
Campus Guana\-jua\-to del Instituto Polit\'ecnico Nacional.\\
Av. Mineral de Valenciana \#200, Col. Fraccionamiento Industrial
Puerto Interior, C.P. 36275, Silao de la Victoria, Guana\-jua\.to,
M\'exico.}

\begin{abstract}In this paper, we present an analysis of a chiral
cosmological scenario from the perspective of K-essence formalism.
In this setup, several scalar fields interact within the kinetic and
potential sectors. However, we only consider a flat
Friedmann--Robertson--Lama\^{\i}tre--Walker universe coupled
minimally to two quintom fields: one quintessence and one phantom.
We examine a classical cosmological framework, where analytical
solutions are obtained. Indeed, we present an explanation of the
``big-bang'' singularity by means of a ``big-bounce''. Moreover,
having a barotropic fluid description and for a particular set of
parameters, the phantom line is in fact crossed. Additionally, for
the quantum counterpart, the Wheeler--DeWitt equation is
analytically solved for various instances, where the factor-ordering
problem has been taken into account (measured by the factor Q).
Hence, this approach allows us to compute the probability density of
the previous two classical subcases. It turns out that its behavior
is in effect damped as the scale factor and the scalar fields
evolve. It also tends towards the phantom sector when the factor
ordering constant $\rm Q\ll 0$.


\end{abstract}
\maketitle
\section{Introduction} 
Over the past decades, various cosmological surveys have suggested
that two stages of accelerated expansion have occurred during the
evolution of the universe~\cite{Perlmutter:1998np, Riess:1998cb,
SupernovaSearchTeam:1997sck, WMAP:2008lyn, Guth, Linde}. The~first
of these epochs, the~so-called inflation~\cite{Guth, Linde}, would
have happened in a very early stage of the expansion of the cosmos,
whilst the second one would be taking place at late times.
Additionally, the consensus is that this accelerated expansion is
caused by dark energy (DE) \cite{Copeland:2006wr, Clifton:2011jh,
Nojiri:2017ncd}. To~account for these phenomena, several
cosmological frameworks incorporate scalar fields into their
prescriptions and, in~fact, they play a preponderant role. Moreover,
one of the most studied scenarios in the literature is the
quintessence model, which is a fluctuating, homogeneous scalar field
that rolls down its scalar potential~\cite{Urena-Lopez:2000ewq,
Ratra:1987rm, Harko:2013gha, Rubano:2001xi, Sahni:1999gb,
Sahni:1999qe, Paliathanasis:2015gga, Dimakis:2016mip}. Different
avenues have been explored, broadening the spectrum of scalar field
models. For~instance, the relevant proposals are the phantom
~\cite{Fang:2004qj, Cataldo:2013de, Nojiri:2015fia},
 quintom~\cite{Cai:2009zp, Setare:2008dw, Lazkoz:2007mx, Leon:2018lnd, Dimakis:2020tzc, Elizalde:2008yf},
 and Chiral fields
~\cite{chervon1995, Chervon:2013btx, Christodoulidis:2019jsx,
Beesham:2013rya, Chervon:2015jji, Fomin:2017bjb, Fomin:2018kui,
Paliathanasis:2018vru}, and there are many
more~\cite{Scherrer:2004au, Bandyopadhyay:2011dh,
Armendariz-Picon:1999hyi, Damour:1992we, Horndeski:1974wa,
Deffayet:2009wt, Coley:1999mj}.

However, despite many efforts
~\cite{Copeland:2006wr,Peebles:2002gy,Padmanabhan:2002ji,Albrecht:2006um,Linder:2008pp,Frieman:2008sn,Caldwell:2009ix},
 the nature of dark energy has not yet been deciphered, except~for its negative pressure. Accordingly,
 the main characteristic of DE is given by its equation of state (EoS), defined by the ratio of the pressure-to
-energy density, that is, $\omega_{DE}\equiv P_{DE}/\rho_{DE}$. This
definition allows us to classify the cosmological
  models mentioned above, according to the behavior of the EoS, namely, quintessence $w_{Q}\ge -1$
  \cite{Ratra:1987rm,Wetterich:1987fm}; phantom $w_{P}\le -1$ \cite{Caldwell:1999ew,Caldwell:2003vq};
  and quintom~\cite{Feng:2004ad}, where the latter is able to evolve across the cosmological
  constant boundary. In~\cite{Cai:2009zp}, the authors have shown that a single scalar field model does not reproduce the
  quintom scenario, thus opening a window to new paradigms where additional degrees of freedom can be
  considered (for non conventional approaches into this matter, we refer the reader to~\cite{Vikman:2004dc, Deffayet:2010qz}).

Our aim is to study a quintom cosmological model. The~most basic
construction of a quintom model can be achieved by considering a
pair of scalar fields, namely, a~canonical one and a phantom one,
endowed with their respective scalar potentials; within this line of
research, different schemes have been considered
~\cite{Cai:2009zp,Setare:2008dw,Lazkoz:2007mx,Leon:2018lnd,Dimakis:2020tzc,
Elizalde:2008yf}.
 These multi-scalar components bring us additional degrees of freedom; thus, various physical phenomena
 can be addressed such as primordial, hybrid~\cite{Chimento:2008ws,lindle,cope,kim,omar-epjp2017}, or
 assisted inflation~\cite{Liddle:1998jc,Copeland:1999cs}, as well as perturbations analysis~\cite{Yokoyama:2007dw,Chiba:2008rp}.

In this work, we present an analysis of a chiral cosmological
scenario from the perspective of K-essence formalism (following the
scheme presented in~\cite{Socorro:2014ama}). In~this prescription,
scalar fields interact within the kinetic and potential sectors. We
consider a Friedmann--Robertson--Lama\^{\i}tre--Walker (FRLW)
universe coupled minimally to two quintom fields: a quintessence and
a phantom. We examine a classical cosmological framework, where
exact solutions are obtained. In~fact, some of them may indicate
that the cosmological singularity is resolved via a ``big-bounce''.
Moreover, we show that the phantom line is crossed. Lastly, for~the
quantum counterpart, the Wheeler--DeWitt (WDW) equation is obtained,
where the factor-ordering problem takes into account the
introduction of the parameter $\rm Q$, and analytical solutions are
presented employing the same relevant cases that appear in the
classical scheme. We show that the probability density is in fact
damped as the scale factor and the scalar fields~evolve.

The paper is laid out as follows. Section~\ref{section2} is devoted
to the analysis of the classical multi-scalar field cosmological
model, and analytical solutions are obtained considering different
cases. In~Section~\ref{section3}, the quantum counterpart is
addressed; in~this formalism, different cases are analyzed and their
corresponding solutions are presented. Section~\ref{section4} is
devoted to the final~remarks.

\section{Classical~Approach}\label{section2}
We start by considering the action of the chiral cosmological model
from the K-essence perspective, which reads
\begin{equation}
S=\int \sqrt{-g}\left[ R - M^{ab}(\phi_c){\cal
G}(\xi_{ab})+C(\phi_c)\right]d^4x, \label{accion}
\end{equation}
where R is the Ricci scalar; $M^{ab}(\phi_c)$ is a matrix related to
the kinetic energy mixed terms; $ C(\phi_c)$ is the scalar
potential, which depends on k scalar fields ($c=1,2,\cdots, k$);
and~${\cal G}[\xi_{ab}(\phi_c)]$ is a functional in terms of the
chiral kinetic energy $\xi_{ab}(\phi_c,g^{\mu
\nu})=-\frac{1}{2}g^{\mu \nu}\nabla_\mu\phi_a \nabla_\nu \phi_b$.
Note that we are working with the reduced Planck units since $8\pi
G=1$, so this eliminates the $8\pi G$ term from the expression
(\ref{accion}). An~action similar to (\ref{accion}) also appears in
modified theories of gravity~\cite{Chervon:2019nwq}, and~more
recently in~\cite{Fomin:2021snm}. Making the variation of the action
(\ref{accion}) with respect to the fields $(g^{\mu \nu}, \phi_c)$,
we obtain
\begin{align}\label{eqn2}
\delta S &= \int {\delta \left[\sqrt{-g}R\right]}d^4x+\int {\delta
\sqrt{-g}}\left[-
M^{ab}(\phi_c){\cal G}(\xi_{ab})+C(\phi_c)\right]d^4x \nonumber\\
& \quad +\int \sqrt{-g}\left[ -{\delta M^{ab}(\phi_c)}{\cal
G}(\xi_{ab})- M^{ab}(\phi_c){\delta
{\cal G}(\xi_{ab})}+{\delta C(\phi_c)}\right]d^4x \,,\nonumber\\
&= \int \sqrt{-g}\, G_{\lambda \theta} { \delta g^{\lambda \theta}}
d^4x - \int \sqrt{-g}\frac{1}{2}\left[- M^{ab}(\phi){\cal
G}(\xi_{ab})+C(\phi_c)\right]g_{\lambda \theta}{\delta g^{\lambda
\theta}} d^4x
+\nonumber\\
&  \quad+\int \sqrt{-g}\left[ {-\frac{\partial
M^{ab}(\phi_c)}{\partial \phi_c}{\delta \phi_c}}{\cal G}(\xi_{ab})-
M^{ab}(\phi_c){\frac{\partial {\cal G}(\xi_{ab})}{\partial
\xi_{ab}}\delta \xi_{ab}}+{\frac{\partial C(\phi_c)}{\partial
\phi}\delta \phi_c}\right]d^4x \,,
\end{align}
where $G_{\lambda \theta}=R_{\lambda \theta}-Rg_{\lambda \theta}/2$,
$R_{\lambda \theta}$, and~$g_{\lambda \theta}$ are the Einstein,
Ricci, and~metric tensors, respectively. The~variation of the
functional  ${\cal G}(\xi_{ab})$ is
\begin{align}
\delta \xi_{ab}(\phi_c)&=\ -\frac{1}{2}\nabla_\mu \phi_a \nabla_\nu
\phi_b{\delta g^{\mu \nu}} - \frac{1}{2}g^{\mu \nu}\nabla_\mu
{\delta \phi_a} \nabla_\nu \phi_b - \frac{1}{2}g^{\mu \nu}\nabla_\mu
\phi_a
\nabla_\nu {\delta \phi_b},\nonumber\\
&= -\frac{1}{2}\nabla_\mu \phi_a \nabla_\nu \phi_b{\delta g^{\mu
\nu}} + \frac{1}{2}\nabla_\mu\left[g^{\mu \nu}  \nabla_\nu
\phi_b\right]{\delta \phi_a}  +
\frac{1}{2}\nabla_\nu\left[g^{\mu \nu}\nabla_\mu \phi_a\right] {\delta \phi_b} \nonumber\\
& \quad +\nabla_\mu\left[\frac{1}{2}\delta \phi_a g^{\mu \nu}
\nabla_\nu \phi_b \right]+\nabla_\nu\left[\frac{1}{2}\delta \phi_b
g^{\mu \nu} \nabla_\nu \phi_a \right] \,.\nonumber
\end{align}

Thus, finally we have
\begin{align}
\delta S &=  \int \sqrt{-g}\left\{G_{\mu
\nu}+\frac{1}{2}M^{ab}(\phi_c) \left[\nabla_\mu \phi_a \nabla_\nu
\phi_b \frac{\partial {\cal G}(\xi_{ab})}{\partial \xi_{ab}}+ g_{\mu
\nu}{\cal G}(\xi_{ab})\right] \right. \nonumber\\
& \left. \quad -\frac{1}{2}g_{\mu \nu} C(\phi_c) \right\} {\delta
g^{\mu \nu}} d^4x \,, \nonumber
\end{align}
and since $\delta\mathcal{S}$ vanishes ($\delta\mathcal{S}=0$) for
arbitrary variations $\delta g^{\mu \nu}$, we are led to the field
equations
\begin{equation}
G_{\mu \nu}=-\frac{1}{2}M^{ab}(\phi_c) \left[\nabla_\mu \phi_a
\nabla_\nu \phi_b \frac{\partial {\cal G}(\xi_{ab})}{\partial
\xi_{ab}}+ g_{\mu \nu}{\cal G}(\xi_{ab})\right] +\frac{1}{2}g_{\mu
\nu} C(\phi_c) \,. \label{EFE}
\end{equation}

Then, the~energy-momentum tensor in this setup becomes
\begin{equation}
 T_{\mu \nu}(\phi_c)=+\frac{1}{2}M^{ab}(\phi_c) \left[\nabla_\mu \phi_a \nabla_\nu \phi_b\frac{\partial {\cal G}(\xi_{ab})}{\partial \xi_{ab}}+ g_{\mu\nu}{\cal G}(\xi_{ab})\right] -\frac{1}{2}g_{\mu \nu} C(\phi_c).
\end{equation}

Moreover, we consider the energy-momentum tensor of a barotropic
perfect fluid $T_{\alpha \beta}(\phi_c)=(\rho + P)u_\alpha(\phi_c)
u_\beta(\phi_c) + P\, g_{\alpha \beta}$ (where the four-velocity is
given by $u_\alpha u_\beta=\nabla_\alpha \phi_a \nabla_\beta
\phi_b/2\xi_{ab}$). Hence, the~pressure $P$ and the energy density
$\rho$ of the scalar fields take the following form:
\begin{equation} \label{presionydensidad}
P(\phi_c)=\frac{1}{2}M^{ab}(\phi_c) {\cal G} -\frac{1}{2}C(\phi_c)
\,, \qquad \rho(\phi_c)=\frac{1}{2}M^{ab}\left[ 2\xi_{ab}
\frac{\partial {\cal G}}{\partial \xi_{ab}}-{\cal G} \right] +
\frac{1}{2}C(\phi_c) \,.
\end{equation}

Additionally, the barotropic parameter $\omega_{\xi_{ab}}$ becomes
\begin{equation}
 \omega_{\xi_{ab}}=\frac{ P(\phi_c)}{
\rho(\phi_c)}=\frac{M^{ab}(\phi_c) {\cal G}-C(\phi_c)}{M^{ab}\left[
2\xi_{ab} \frac{\partial {\cal G}}{\partial \xi_{ab}}-{\cal G}
\right] + C(\phi_c)} \,. \label{barotro}
\end{equation}

On the other hand, taking the variation of the action (\ref{accion})
with respect to the scalar field $\phi_c$, we obtain
\begin{eqnarray}
&& \delta S = \int \sqrt{g} \left\{-\frac{\partial
M^{ab}(\phi_c)}{\partial \phi_c} {\cal G}(\xi_{ab})
-\frac{1}{2}M^{cb}(\phi_c)\frac{\partial {\cal
G}(\xi_{ab})}{\partial \xi_{ab}}\nabla_\nu \nabla^\nu \phi_b \right. \nonumber\\
&& \left. \qquad- \frac{1}{2}M^{ac}(\phi_c)\frac{\partial {\cal
G}(\xi_{ab})}{\partial \xi_{ab}}\nabla_\nu \nabla^\nu \phi_a
+\frac{\partial C(\phi_c)}{\partial \phi_c} \right\} {\delta \phi_c}
d^4 x \,, \nonumber
\end{eqnarray}
where a Klein-Gordon-like equation can be written as follows:
\begin{equation}\label{eqn8}
\frac{\partial M^{ab}(\phi_c)}{\partial \phi_c} {\cal G}(\xi_{ab})
-M^{ac}(\phi_c)\frac{\partial {\cal G}(\xi_{ab})}{\partial
\xi_{ab}}\nabla_\nu \nabla^\nu \phi_a -\frac{\partial
C(\phi_c)}{\partial \phi_c}=0 \,.
\end{equation}

Note that Equations~(\ref{eqn2})-(\ref{eqn8}) represent the general
framework; however, we will particularize to the case $ {\cal
G}=\xi$, therefore obtaining the standard chiral Einstein field
equations
\begin{equation}
G_{\mu \nu}=-\frac{1}{2}M^{ab}(\phi_c) \left[\nabla_\mu \phi_a
\nabla_\nu \phi_b - \frac{1}{2}g_{\mu \nu} g^{\alpha
\beta}\nabla_\alpha \phi_a \nabla_\beta \phi_b \right]
+\frac{1}{2}g_{\mu \nu} C(\phi_c) \,. \label{einstein}
\end{equation}

Now, if~we consider that $M^{ab}$ is a constant matrix, we obtain
\begin{equation}
M^{ac}\nabla_\nu \nabla^\nu \phi_a -\frac{\partial
C(\phi_c)}{\partial \phi_c}=0 \,. \label{K-G}
\end{equation}

All of the aforementioned results can be employed to consider a
two-field cosmological model: a {\bf quin}tessence and a phan{\bf
tom} field, with their
corresponding scalar potentials. Setting $M^{ab}(\phi_c)=m^{ab}$ as
a constant matrix, in~(\ref{accion}), we obtain
\begin{equation}
{\cal L}=\sqrt{-g} \left( R-\frac{1}{2}g^{\mu\nu} m^{ab}\nabla_\mu
\phi_a \nabla_\nu \phi_b  + V(\phi_1,\phi_2)\right) \,,
\label{lagra}
\end{equation}
where $ C(\phi_c)=V(\phi_1,\phi_2)$ is the combined scalar field
potential; $\phi_1$ and $\phi_2$ are the quintessence and phantom
fields, respectively; and $ m^{ab}$ is a
 $2 \times 2$ constant matrix of the form
$m^{ab}=\left(
\begin{tabular}{cc}
$1$ & $ m^{12}$ \\
$ m^{12}$ & $ -1$ \end{tabular}
 \right). $

Thus, the~Einstein--Klein--Gordon field Equations~(\ref{einstein})
and (\ref{K-G}) are
\begin{equation}
G_{\alpha \beta}= -\frac{1}{2}m^{ab} \left(\nabla_\alpha \phi_a
\nabla_\beta \phi_b -\frac{1}{2}g_{\alpha \beta} g^{\mu \nu}
\nabla_\mu \phi_a \nabla_\nu \phi_b \right) +\frac{1}{2}g_{\alpha
\beta} \, V(\phi_1,\phi_2) \,, \label{mono}
\end{equation}
\begin{equation}
m^{cb} \nabla_\nu \nabla^\nu \phi_b -\frac{\partial
C(\phi_c)}{\partial \phi_c}=0 \,,
\end{equation}
where $a,b,c=1,2$. From~(\ref{mono}), the~energy-momentum tensor of
the scalar fields $ (\phi_1,\phi_2)$ is given by
\begin{equation}
 8\pi G T_{\alpha \beta}(\phi_1,\phi_2)=-\frac{1}{2}m^{ab}
\left(\nabla_\alpha \phi_a \nabla_\beta \phi_b -\frac{1}{2}g_{\alpha
\beta} g^{\mu \nu} \nabla_\mu \phi_a \nabla_\nu \phi_b \right)
+\frac{1}{2}g_{\alpha \beta} \, V(\phi_1,\phi_2) \,, \label{tmunu}
\end{equation}
then, using (\ref{barotro}), the~barotropic index
$\omega_{\phi_a,\phi_b}$ is given by
\begin{equation}
\omega_{\phi_a\,\phi_b}=\frac{-\frac{1}{2}m^{ab}\nabla^\mu \phi_a
\nabla_\mu \phi_b-C(\phi_c)}{-\frac{1}{2}m^{ab}\nabla^\mu \phi_a
\nabla_\mu \phi_b + C(\phi_c)} \,.
\end{equation}

In our analysis, the background spacetime to be considered is a
spatially flat FRLW with line element
\begin{equation}
ds^2=-N(t)^2 dt^2 +e^{2\Omega(t)} \left[dr^2
+r^2(d\theta^2+sin^2\theta d\phi^2) \right] \,, \label{frw}
\end{equation}
where $N$ represents the lapse function, $ A(t)=e^{\Omega(t)}$ is
the scale factor in the Misner parametrization, and  $\Omega$ is a
scalar function whose interval is $ (-\infty,\infty)$. Choosing $
C(\phi_c)=V_1(\phi_1)+V_2(\phi_2)=V_{01}e^{-\lambda_1\phi_1}+V_{02}e^{-\lambda_2
\phi_2}$, the~mixed Einstein field equations are

\begin{align}
 3\frac{\dot
\Omega^2}{N^2}-\frac{1}{2}\left[\frac{1}{2}\frac{\dot
\phi_1^2}{N^2}+V_1(\phi_1)\right]-\frac{1}{2}\left[-\frac{1}{2}\frac{\dot
\phi_2^2}{N^2}+V_2(\phi_2) \right]-\frac{m^{12}}{2}\frac{\dot
\phi_1}{N} \frac{\dot \phi_2}{N}&=\rm 0 \,, \label{ein0}\\
 2\frac{\ddot \Omega}{N^2}+3\frac{\dot \Omega^2}{N^2}-2\frac{\dot
\Omega}{N}\frac{\dot N}{N^2}+ \frac{1}{2}\left[\frac{1}{2}\frac{\dot
\phi_1^2}{N^2}-V_1(\phi_1)
\right]+\frac{1}{2}\left[-\frac{1}{2}\frac{\dot
\phi_2^2}{N^2}-V_2(\phi_2) \right]+\frac{m^{12}}{2}\frac{\dot
\phi_1}{N} \frac{\dot \phi_2}{N}&=\rm 0 \,, \label{ein1}\\
 \dot \phi_1\left(-3\frac{\dot \Omega}{N}\frac{\dot
\phi_1}{N}+\frac{\dot N}{N^2}\frac{\do \phi_1}{N}-\frac{\ddot
\phi_1}{N^2} \right)+ m^{12}\dot \phi_1\left(-3\frac{\dot
\Omega}{N}\frac{\dot \phi_2}{N}+\frac{\dot N}{N^2}\frac{\do
\phi_2}{N}-\frac{\ddot \phi_2}{N^2} \right)-\dot
V_1(\phi_1)&=\rm 0 \,,\label{kg1}\\
m^{12}\dot \phi_2\left(-3\frac{\dot \Omega}{N}\frac{\dot
\phi_1}{N}+\frac{\dot N}{N^2}\frac{\do \phi_1}{N}-\frac{\ddot
\phi_1}{N^2} \right)+ \dot \phi_2\left(3\frac{\dot
\Omega}{N}\frac{\dot \phi_2}{N}-\frac{\dot N}{N^2}\frac{\do
\phi_2}{N}+\frac{\ddot \phi_2}{N^2} \right)-\dot V_2(\phi_2)&=\rm 0
\,,\label{kg2}
\end{align}
where $``\cdot"$ represents a time derivative. By plugging the line
element (\ref{frw}) into the energy-momentum tensor of the scalar
fields (\ref{tmunu}), the~energy density, and the pressure, the
following form is taken 
:
\begin{align}
8\pi G \rho_{\phi_1 \phi_2}&= \frac{1}{2}\left[\frac{1}{2}\dot
\phi_1^2+N^2 V_1(\phi_1)\right]+\frac{1}{2}\left[-\frac{1}{2}\dot
\phi_2^2+N^2 V_2(\phi_2) \right]+\frac{m^{12}}{2}\dot
\phi_1 \dot \phi_2 \,,\label{rho}\\
8\pi G P_{\phi_1 \phi_2}&= \frac{1}{2}\left[\frac{1}{2}\frac{\dot
\phi_1^2}{N^2}-V_1(\phi_1)
\right]+\frac{1}{2}\left[-\frac{1}{2}\frac{\dot
\phi_2^2}{N^2}-V_2(\phi_2) \right]+\frac{m^{12}}{2}\frac{\dot
\phi_1}{N} \frac{\dot \phi_2}{N} \,, \label{presion}
\end{align}
having these two quantities at hand, the barotropic parameter will
be written as
\begin{equation}
\omega_{\phi_1 \phi_2}= \frac{P_{\phi_1
\phi_2}}{\rho_{\phi_1\phi_2}}= \frac{\left[\frac{1}{2}\frac{\dot
\phi_1^2}{N^2}-V_1(\phi_1) \right]+\left[-\frac{1}{2}\frac{\dot
\phi_2^2}{N^2}-V_2(\phi_2) \right]+m^{12}\frac{\dot \phi_1}{N}
\frac{\dot \phi_2}{N}}{\left[\frac{1}{2}\dot \phi_1^2+N^2
V_1(\phi_1)\right]+\left[-\frac{1}{2}\dot \phi_2^2+N^2 V_2(\phi_2)
\right]+m^{12}\dot \phi_1 \dot \phi_2} \,. \label{omega-parameter}
\end{equation}

Now we are in position to construct the corresponding Lagrangian and
Hamiltonian densities for this cosmological model. Using Hamilton's
approach, classical solutions to EKG~(\ref{ein0})--(\ref{kg2}) can
be found; additionally, the quantum counterpart can be established
and solved. Taking these ideas into consideration, putting back the
metric (\ref{frw}) into (\ref{lagra}), the Lagrangian density reads
\begin{equation}\label{lagrafrw}
{\cal{L}}=
e^{3\Omega}\left(\frac{6\dot{\Omega}^2}{N}-\frac{\dot{\phi_1}^2}{2N}+\frac{\dot{\phi_2}^2}{2N}
-\frac{m^{12}\dot{\phi_1} \dot \phi_2}{N}+ N V_{01} e^{-\lambda_1
\phi_1} +N V_{02} e^{-\lambda_2 \phi_2} \right)\,.
\end{equation}

The resulting momenta are given by
\begin{equation}\label{mom_conjugados_26}
\begin{split}
\Pi_\Omega &= 12 \frac{e^{3\Omega}}{N}\dot \Omega \,,\\
\Pi_{\phi_1}&=  -\frac{e^{3\Omega}}{N}\left(\dot\phi_1 +m^{12} \dot \phi_2\right) \,,\\
\Pi_{\phi_2}&= -\frac{e^{3\Omega}}{N}\left( m^{12}\dot\phi_1 - \dot
\phi_2\right) \,,
\end{split}
\qquad
\begin{split}
\dot \Omega&= \frac{N e^{-3\Omega}}{12} \Pi_\Omega \,,\\
 \dot \phi_1&=-\frac{N e^{-3\Omega}}{\triangle}\left(\Pi_{\phi_1}+m^{12} \Pi_{\phi_2} \right),\\
  \phi_2&= \frac{N e^{-3\Omega}}{\triangle}\left(
m^{12}\Pi_{\phi_1}-\Pi_{\phi_2} \right) \,,
\end{split}
\end{equation}
where $\triangle =1+\left(m^{12}\right)^2$. In~order to obtain a
Hamiltonian density, we write (\ref{lagrafrw}) in a canonical form,
i.e.,~$\mathcal L_{can}=\Pi_q\dot q-N\mathcal H$; then, we perform
the variation with respect to the lapse function $N$,
$\delta\mathcal L_{can}/\delta N=0$, yielding the Hamiltonian
constraint $\mathcal H=0$, that is,
\vspace{-6pt}

\begin{equation}
{\cal H}=  \frac{e^{-3\Omega}}{24} \left[
\Pi_\Omega^2-\frac{12}{\triangle}\Pi_{\phi_1}^2+
\frac{12}{\triangle}\Pi_{\phi_2}^2-24\frac{m^{12}}{\triangle}\Pi_{\phi_1}\Pi_{\phi_2}
-24V_1 e^{-\lambda_1\phi_1+6\Omega} -24V_2
e^{-\lambda_2\phi_2+6\Omega}\right] \,. \label{hamifrw}
\end{equation}

The fact that ${\cal H}=0$ guarantees us that its solutions are
unique and well defined. Putting forward the following canonical
transformation on the variables
$(\Omega,\phi_1,\phi_2)\leftrightarrow (\xi_1,\xi_2,\xi_3)$
and~fixing the gauge $N=24e^{3\Omega}$, we obtain
\begin{equation}\label{trans_2}
\begin{split}
\xi_1&=6\Omega-\lambda_1 \phi_1 \,,\\
\xi_2&= 6 \Omega-\lambda_2 \phi_2 \,,\\
 \xi_3&=6 \Omega + \lambda_1
\phi_1 + \lambda_2\phi_2 \,,
\end{split}
\quad\longleftrightarrow\quad
\begin{split}
\Omega&=\frac{\xi_1 + \xi_2+  \xi_3}{18} \,,\\
\phi_1&= \frac{-2\xi_1 + \xi_2+\xi_3}{3\lambda_1} \,,\\
\phi_2 &= \frac{\xi_1-2\xi_2+\xi_3}{3\lambda_2} \,,
\end{split}
\end{equation}
leading us to obtain a new set of conjugate momenta $(P_1,P_2,P_3)$
\begin{align}
\Pi_\Omega &= 6 P_1 +6 P_2+ 6 P_3 \,, \nonumber\\
\Pi_{\phi_1} &= \lambda_1 \left(-P_1 + P_3\right) \,,
\nonumber\\
\Pi_{\phi_2} &= \lambda_2\left(- P_2 + P_3\right) \,,
\label{new-moment}
\end{align}
therefore, the Hamiltonian density can be written as
\begin{align}
{\cal H} &= 12 \left(3-\Lambda_1
 \right)P_1^2+12\left(3+\Lambda_2
\right)P_2^2+ 12\left(3 - 2\Lambda_{12} +
\Lambda_2-\Lambda_1\right)P_3^2 \nonumber\\
&\quad + 24\left[ \left(3+ \Lambda_1+ \Lambda_{12}\right)P_1+
\left(3 + \Lambda_{12} -\Lambda_2\right)P_2\right]
P_3 \nonumber\\
&\quad +24\left(3-\Lambda_{12}\right)P_1 P_2- 24\left(V_1
e^{\xi_1}+V_2 e^{\xi_2}\right) \,,\label{hamifrw_new}
\end{align}
where $\Lambda_1=\lambda_1^2/\triangle$,
$\Lambda_2=\lambda_2^2/\triangle$, and
$\Lambda_{12}=m^{12}\,\lambda_1 \lambda_2/\triangle$. In~the end,
even if the Hamiltonian density (\ref{hamifrw_new}) exhibits an
intricate form, this configuration will indeed allow us to compute
various relevant scenarios. Thus, the Hamilton equations become
\begin{align}
\dot \xi_1&= 24 \left(3- \Lambda_{1}\right)P_1
+24\left(3-\Lambda_{12} \right)P_2+ 24 \left(3+
\Lambda_1+\Lambda_{12}
\right)P_3 \,,\nonumber\\
\dot \xi_2&= 24\left(3+ \Lambda_2\right)P_2 +24\left(3-
\Lambda_{12}\right)P_1+ 24\left(3-   \Lambda_2+
\Lambda_{12} \right)P_3 \,, \nonumber\\
\dot \xi_3 &= 24\left(3+\Lambda_1+\Lambda_{12}\right)P_1+
24\left(3 - \Lambda_2+ \Lambda_{12}\right)P_2 +24 \left(3+\Lambda_2-\Lambda_1-2\Lambda_{12}\right)P_3 \,,\nonumber\\
 \dot P_1 &=  24 V_1e^{\xi_1},\label{ecs_mov_2}\\
 \dot P_2 &= 24V_2e^{\xi_2} \,,\nonumber \\
 \dot P_3 &= 0 \,. \nonumber
\end{align}

Right away, we can see that $P_3=p_3=constant$. Moreover, the~end game of this analysis is to find solutions to the variables ($\Omega, \phi_1, \phi_2$) 
. Hence, we simplify our expression. First, we drop the mixed
momenta $\rm P_1$ and $\rm P_2$ from $\dot{\xi}_{1}$ and
$\dot{\xi}_{2}$ (Equation~(\ref{ecs_mov_2})) by~setting their
coefficients to zero: $3-\Lambda_{12}=0$. Therefore, we can obtain
one relation among the parameters $(m^{12},\lambda_1\lambda_2)$,
where the matrix element $m^{12}$ satisfies the constraint
\begin{equation}
m^{12}=\frac{\lambda_1 \lambda_2}{6}\left[1 \pm \sqrt{1- \left(
\frac{6}{\lambda_1 \lambda_2}\right)^2}\right] \,.
\label{constraint}
\end{equation}

Additionally, we set the second term inside the square root of
(\ref{constraint}) to be a real number and consider $\lambda_1>0$,
$\lambda_2>0$, thus yielding the relation $\lambda_1\lambda_2 \geq
6$, ensuring that $m^{12}$ is always~positive.
%
%
\subsection{Classical Exact~Solutions}

 In this section, we will calculate the exact solutions of ($\Omega, \phi_1, \phi_2$), where different cases
will appear due to the parameters $(\lambda_1,\lambda_2)$. Recall
the master Hamiltonian density
\begin{align}
{\cal H} &=  12 \eta_1 P_1^2+12\eta_2P_2^2+ 12\left(-9 +
\eta_1+\eta_2\right)p_3^2  + 24\left[ \left(9-\eta_1 \right)P_1+
\left(9 -\eta_2\right)P_2\right] p_3 \nonumber\\
& - 24\left(V_1 e^{\xi_1}+V_2 e^{\xi_2}\right)
\,,\label{hamiltonian}
\end{align}
with $\eta_1=3-\Lambda_1$ and $\eta_2=3+\Lambda_2$. Then, Hamilton
equations for these new coordinates $\xi_i$ are
\begin{eqnarray}\label{new_ecs_mov_2}
\dot \xi_1&=& 24 \eta_1P_1 + 24 \left(9-\eta_1
\right)p_3 \,,\nonumber\\
\dot \xi_2&=& 24\eta_2P_2 + 24\left(9-\eta_2 \right)p_3 \,, \\
\dot \xi_3 &=&24\left(9-\eta_1\right)P_1+ 24\left(9
-\eta_2\right)P_2 +24 \left(-9+\eta_1+\eta_2\right)p_3 \,,\nonumber
\end{eqnarray}
and equations for $\dot P_1$ and $\dot P_2$ are still given by
(\ref{ecs_mov_2}). In~the following sections, we will obtain
analytical solutions for differents values of $\lambda_1$ and
$\lambda_2$.
%
%
\subsubsection{Case: $\lambda_1=\lambda_2=\sqrt{6}$.}
For these particular values, we have $\Lambda_1=\Lambda_2=3$ with
$\eta_1=0$ and $\eta_2=6$; then, the Hamilton equations are reduced
to
\begin{eqnarray}\label{34}
\dot \xi_1 &=&  216 p_3 \,, \nonumber\\
\dot \xi_2 &=& 144P_2 + 72p_3 \,, \\
\dot \xi_3 &=&216P_1+72P_2 -72p_3 \,.\nonumber
\end{eqnarray}

From the last set of equations, we can see that the solution for
$\dot \xi_1$ will be given by
\begin{equation}
\xi_1 = a_1+216 p_3 t \,,
\end{equation}
where $a_1$ is an integration constant. Then, taking the time
derivative of $\dot\xi_2$ results in $\ddot \xi_2=3456
V_2\,e^{\xi_2}$, whose solution is
\begin{equation}
\xi_2=Ln\left(\frac{\alpha_2^2}{1728\, V_2}
\right)+Ln\left[Csch^2(\alpha_2\,t-\beta_2) \right].
\end{equation}

Now we know the functional form of $\xi_2$, we can compute the
remaining momenta, yielding
\begin{eqnarray}
P_1(t)&=& p_1+\frac{V_1}{9p_3}\, e^{a_1+216p_3\,t},\nonumber\\
P_2(t)&=& p_2-\frac{\alpha_2}{72}\, Coth(\alpha_2\,t - \beta_2) \,.
\label{37}
\end{eqnarray}

Plugging back $P_1$ and $P_2$, given by (\ref{37}), into~the
Hamiltonian constraint ${\cal H}=0$, we found that $p_2=-p_3/2$ and
$3888p^2_3-15552 p_1 p_3 -\alpha_2^2=0$; solving for $ p_3$ gives
$p_3=2p_1 \pm \frac{\sqrt{3}}{108}\sqrt{\alpha_2^2+15552\,p_1^2}$.

With these results, the variable $\xi_3$ becomes
\begin{equation}
\xi_3=a_3+(216p_1-108p_3)t+\frac{V_1}{9p_3^2}e^{a_1+216p_3t}+Ln\left[Csch\left(\alpha_2\,t-\beta_2\right)
\right] \,.
\end{equation}
where $ a_3$ is an integration constant. Having found $\xi_1$,
$\xi_2$, and~$\xi_3$ and then applying the inverse transformation
(\ref{trans_2}), we can present the solutions in the original
variables
\begin{align}
 \Omega(t)&= \frac{a_1+a_3}{18}+
Ln\left[\frac{\alpha_2}{24\sqrt{3\,V_2}} \right]^{\frac{1}{9}}
\nonumber\\
& \quad + Ln\left[Csch^{\frac{1}{6}}\left(\alpha_2\,t-\beta_2
\right) \right]+(12p_1+6p_3)t
+\frac{V_1}{162p_3^2}e^{a_1+216p_3t} \,,\label{omega}\\
 \phi_1(t)&=\frac{-2a_1+a_3}{3\lambda_1}+Ln\left(\frac{\alpha_2^2}{1728\, V_2}
\right)^{\frac{1}{3\lambda_1}} \nonumber\\
& \quad +\frac{1}{\lambda_1}\left[\left(72p_1-180p_3\right)t+
\frac{V_1}{27p_3^2}e^{a_1+216p_3t}+Ln\left[Csch\left(\alpha_2\,t-\beta_2 \right) \right]\right] \,, \label{phi1}\\
\phi_2(t)&=\frac{a_1+a_3}{3\lambda_2}+Ln\left(\frac{\alpha_2^2}{1728\,
V_2}
\right)^{-\frac{2}{3\lambda_2}} \nonumber\\
& \quad + \frac{1}{\lambda_2} \left[\left(72p_1+36p_3\right)t+
\frac{V_1}{27p_{3}^2}e^{a_1+216p_3t}+Ln\left[Sinh\left(\alpha_2\,t-\beta_2
\right) \right] \right] \,, \label{phi12}
\end{align}

Recalling that the scale factor is given by $A(t)=e^{\Omega(t)}$, we
have
\begin{equation}
A(t)=e^{\frac{a_1+a_3}{18}}\,\left[\frac{\alpha_2}{24\sqrt{3 V_2}}
\right]^{\frac{1}{9}}\, Csch^{\frac{1}{6}}\left(\alpha_2\,t-\beta_2
\right)\,\,
Exp\left[\frac{V_1}{162p_3^2}e^{a_1+216p_3t}\right]e^{(12p_1+6p_3)t}
\,. \label{scalefactor}
\end{equation}

\begin{figure}[h]
\includegraphics[scale=0.3]{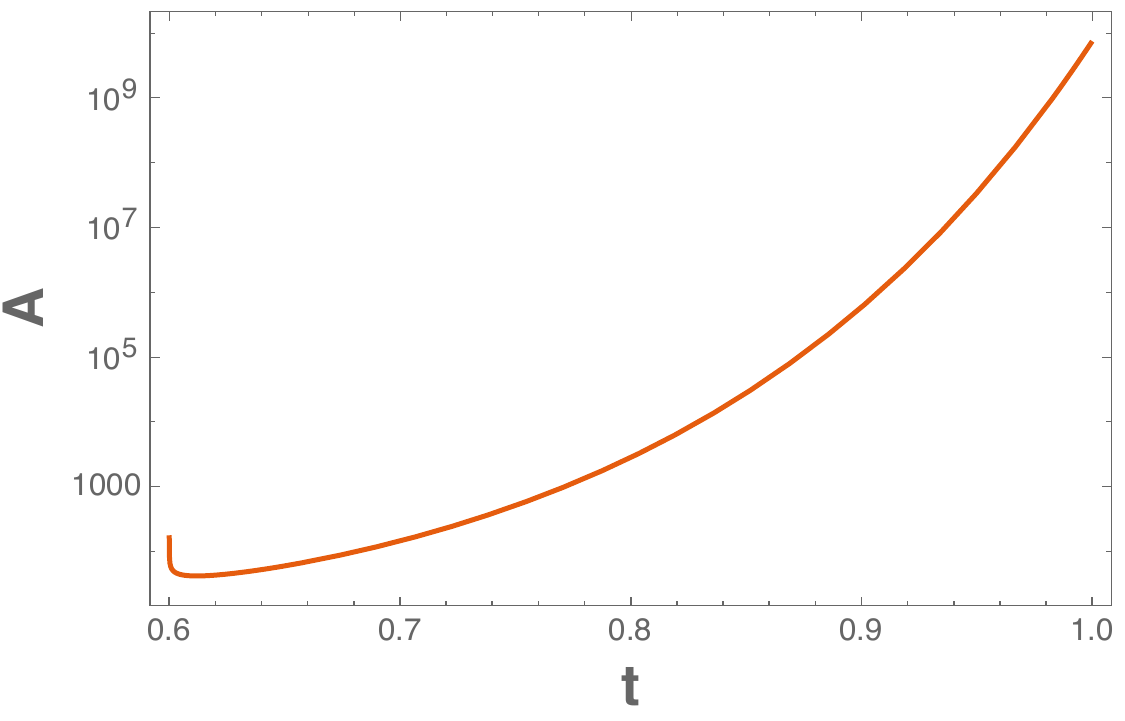}
\includegraphics[scale=0.3]{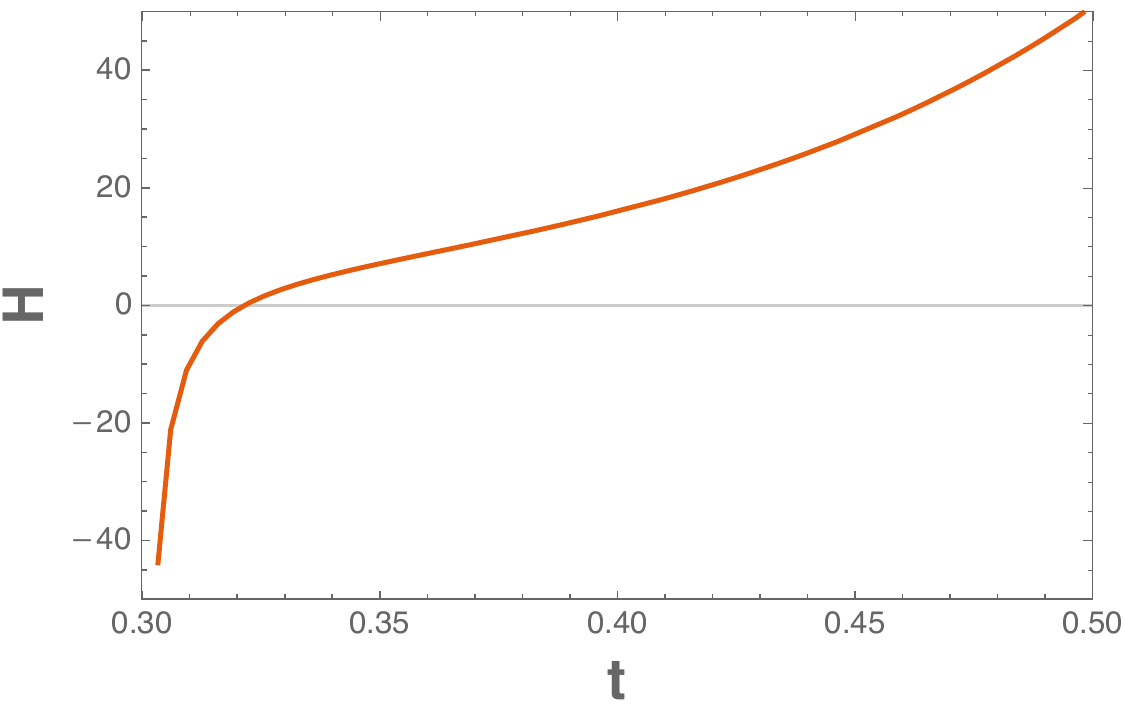}
\includegraphics[scale=0.3]{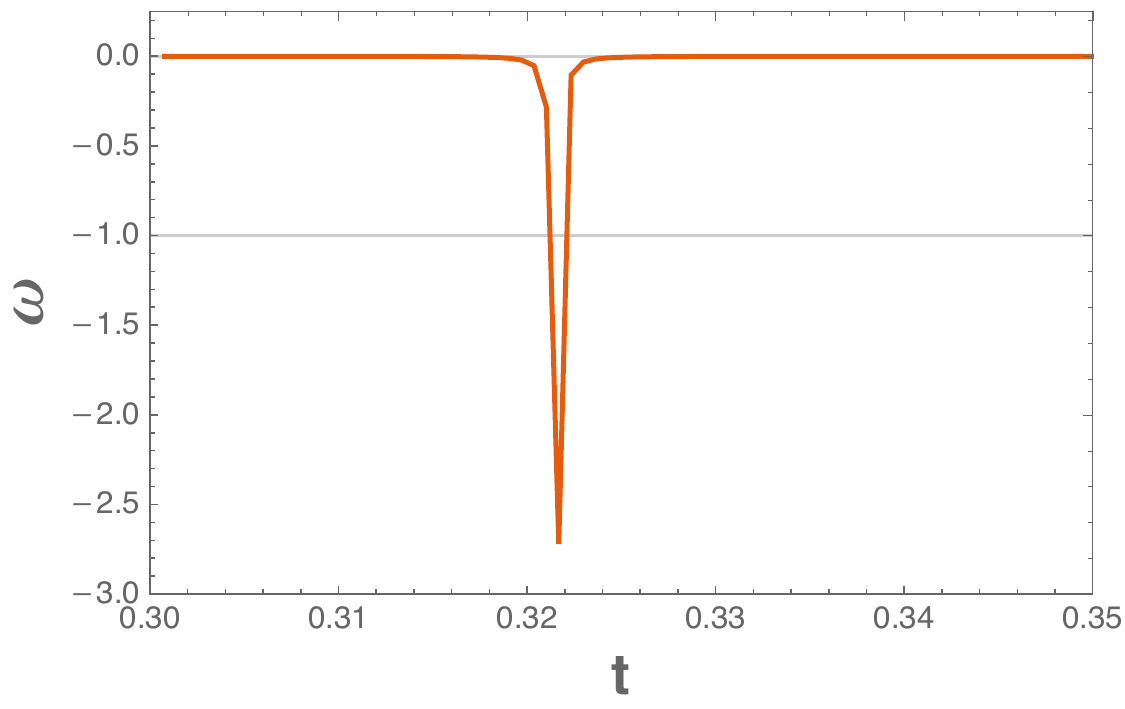}
\caption{
This figure shows the time ($0.3\leq t\leq 1.0$) evolution of the
scale factor $A(t)$, the~Hubble parameter $ H(t)$, and~the
barotropic parameter $\omega_{\phi_1 \phi_2}(t)$. We use arbitrary
units, namely, $ V_1 = 6.0 \,, V_2 = 0.1$, $ \alpha_2=3.0$, $ a_1=
-6.548 \,, a_3=-2.0$, and~$p_{1}=0.001$. Recall that
$\lambda_1=\lambda_2=\sqrt{6}$; the~remaining constants can be
obtained from the aforementioned values. Note that time is measured
in reduced Planck units since $8\pi G=1$. }\label{figura_1}
\end{figure}

In Figure~\ref{figura_1}, we present the behaviour of the scale
factor $A=A(t)$, the~Hubble parameter $H=H(t)$, and~the barotropic
parameter $\omega_{\phi_1 \phi_2}=\omega_{\phi_1 \phi_2}(t)$.
From~the upper left graph, we can see that $A$ grows very rapidly as
time goes by; it can also be seen that this solution avoids the
singularity by means of a bounce, where $H$ does cross the
horizontal axis. In~the panel at the bottom, the~barotropic
parameter $\omega_{\phi_1 \phi_2}$ is presented, and~it can be seen
that the EoS parameter crosses the ``$-$1'' boundary, which is in
fact a characteristic of the quintom~models.

\subsubsection{Case: $\lambda_1\lambda_2=6$.}
Now, we have $m^{12}=1$, and~for $\lambda_2=\sqrt{6}$ we obtain the
previous case; therefore, we devote this section to carrying out an
analysis of values $\lambda_2\neq\sqrt{6}$ and~explore whether the
phantom or quintessence scheme prevails under the domain of the
scalar potential. On~the one hand, when $\lambda_2\ll\lambda_1$ the
phantom sector dominates. On~the other hand, when
$\lambda_2\gg\lambda_1$, the quintessence counterpart becomes the
relevant scenario. Then, we consider the Hamilton
Equations~(\ref{new_ecs_mov_2}) and take the time derivative of
$\dot \xi_1$, which reads
\begin{equation}
\ddot \xi_1= 576\eta_1 \,V_1 e^{\xi_1} \,, \label{xi-1}
\end{equation}
where we also resort to the equation for $\dot P_1$. Solutions of
(\ref{xi-1}) strongly depend on $\lambda_1$, which has the form
\begin{equation}
e^{\xi_1}=\frac{ r_1^2}{288|\eta_1|\,V_1}\left\{
\begin{tabular}{ll}
$Sech^2\left(r_1 t-q_1\right)$ & $\lambda_1 > \sqrt{6}$
\mbox{corresponding at} $\eta_1<0$  \\
$Csch^2\left(r_1 t-q_1\right)$ & $ \lambda_1 < \sqrt{6}$
\mbox{corresponding at} $\eta_1>0$  \label{solucion-xi1}
\end{tabular}
\right.
\end{equation}

From (\ref{new_ecs_mov_2}), we can see that both $\dot\xi_2$ and
$\dot\xi_1$ have the same functional structure when $\eta_1>0$,
and~since $\eta_2>0$ for all values of $\lambda_2$, the~solution of
$\xi_2$ is
\begin{equation}
 e^{\xi_2}=\frac{ r_2^2}{288\eta_2\,V_2} \, Csch^2\left(r_2
t-q_2\right) \,, \label{solucion-xi2}
\end{equation}
where in (\ref{solucion-xi1}) and (\ref{solucion-xi2}), $r_i$ and
$q_i$ (with $i=1,2$) are integration constants. In~the next
segments, we will examine the two cases: $\lambda_1>\sqrt{6}$ and
$\lambda_1<\sqrt{6}$.
%
%
\subsubsection{Phantom Domination: $\lambda_1>\sqrt{6}$ and $\lambda_2<\sqrt{6}$.}
Considering this setup, we start by reinserting the solutions for
$\lambda_1>\sqrt{6}$ ($\eta_1 <0$) and for $\lambda_2<\sqrt{6}$ into
the Hamilton equations for the momenta, obtaining
\begin{eqnarray}
P_1 &=& p_1 + \frac{r_1}{12|\eta_1|} \, Tanh\left(r_1 t-q_1
\right) \,, \label{solution-p1} \\
P_2 &=& p_2 - \frac{r_2}{12\eta_2} \, Coth\left(r_2 t-q_2 \right)
\,, \label{solution-p2}
\end{eqnarray}
where $p_1$ and $p_2$ are integration constants. Now, with~the aid
of Equations~(\ref{solution-p1}) and (\ref{solution-p2}),
the~Hamiltonian is identically zero when
\begin{equation}
p_1=\frac{|\eta_1|+9}{|\eta_1|} p_3 \,, \qquad
p_2=\frac{\eta_2-9}{\eta_2} p_3 \,,\qquad p_3= +\frac{1}{36}
\sqrt{\frac{\eta_2 r_1^2-|\eta_1| r_2^2
}{3\left[|\eta_1|\eta_2-3|\eta_1| +3\eta_2 \right]}} \,.
\label{coefficients}
\end{equation}

Consequently, the~solutions of $\xi_i$ become
\begin{align}
\xi_1 &= Ln\left(\frac{r_1^2}{288|\eta_1|\, V_1}\right)+Ln\left[
Sech^2
\left(r_1\,t -q_1 \right)\right] \,,  \label{xi1-1}\\
\xi_2 &= Ln\left(\frac{r_2^2}{288\eta_2\, V_2}\right)+Ln\left[
Csch^2
\left(r_2\,t -q_2 \right)\right] \,, \label{xi2-1}\\
\xi_3&= a_3 + 648\frac{|\eta_1|\, \eta_2-3|\eta_1| +
3\eta_2}{|\eta_1|\,\eta_2} p_3 t
+\frac{9+|\eta_1|}{|\eta_1|}Ln\left[Cosh^2\left(r_1\,t - q_1 \right)
\right]\nonumber\\
&\quad +\frac{\eta_2-9}{\eta_2}Ln\left[Sinh^2\left(r_2\,t - q_2
\right) \right] \,,
\end{align}
where $a_3$ is an integration constant. To~arrive at the solutions
in terms of the original variables $(\Omega, \phi_1, \phi_2)$, we
apply the inverse canonical transformation (\ref{trans_2}),
obtaining the following:
\begin{eqnarray}
\Omega &=& \Omega_0 + Ln\left[ Cosh^{\beta_1} \left(r_1\,t -q_1
\right) Csch^{\beta_2}\left(r_2\, t -q_2
\right)\right]\nonumber\\
&&\quad +36\frac{|\eta_1|\, \eta_2-3|\eta_1| +
3\eta_2}{|\eta_1|\,\eta_2} p_3 t \,,
\\
\phi_1 &=& \phi_{10}+ Ln\left[Cosh^{\frac{2\left(|\eta_1|+3
\right)}{\lambda_1 |\eta_1|}}\left(r_1\,t-q_1\right)
Csch^{\frac{6}{\lambda_1 \eta_2}}\left(r_2\, t -q_2\right)\right]
\nonumber\\
&& \quad +216\frac{|\eta_1|\, \eta_2-3|\eta_1| +
3\eta_2}{\lambda_1\, |\eta_1|\,\eta_2} p_3 t  \,,\label{phi-1} \\
 \phi_2 &=&
\phi_{20}+Ln\left[Cosh^{\frac{6}{\lambda_2
|\eta_1|}}\left(r_1\,t-q_1\right) Sinh^{\frac{2\left(
\eta_2-3\right)}{\lambda_2 \eta_2}}\left(r_2\, t -q_2\right)\right]
\nonumber\\
 && \quad +216\frac{|\eta_1|\, \eta_2-3|\eta_1| +
3\eta_2}{\lambda_2\, |\eta_1|\,\eta_2} p_3 t  \,,\label{phi-2}
\end{eqnarray}
where $\beta_1=1/|\eta_1|$, $\beta_2=1/\eta_2$, and~the constants
$\Omega_0, \phi_{10}$ and $\phi_{20}$ are given by
\begin{eqnarray}\label{const}
\Omega_0&=&Ln\left[\frac{r_1\,r_2}{288\sqrt{|\eta_1|\eta_2
V_1\,V_2}} \right]^{\frac{1}{9}}+\frac{a_3}{18} \,, \nonumber\\
\phi_{10}&=& Ln\left[\frac{12\sqrt{2}r_2\,|\eta_1|
V_1}{r_1^2\sqrt{\eta_2 V_2}}
 \right]^{\frac{2}{3\lambda_1}}+\frac{a_3}{3\lambda_1} \,,\\
\phi_{20}&=&Ln\left[\frac{12\sqrt{2}r_1\eta_2
V_2}{r_2^2\sqrt{|\eta_1| V_1}}
\right]^{\frac{2}{3\lambda_2}}+\frac{a_3}{3\lambda_2}\nonumber \,.
\end{eqnarray}

For this case, the~scale factor becomes
\begin{align}
A(t)&=\left[\frac{r_1 r_2}{288\sqrt{|\eta_1|\eta_2 V_1 V_2}}
\right]^{\frac{1}{9}}\,e^{\frac{a_3}{18}} Cosh^{\beta_1}
\left(r_1\,t -q_1 \right) Csch^{\beta_2}\left(r_2\, t -q_2
\right)\,\nonumber\\
&\quad \times Exp\left[36\frac{|\eta_1|\, \eta_2-3|\eta_1| +
3\eta_2}{|\eta_1|\,\eta_2} p_3 t\right]
\,.\label{scale-factor-phantom}
\end{align}

\begin{figure}[h]
\includegraphics[scale=0.3]{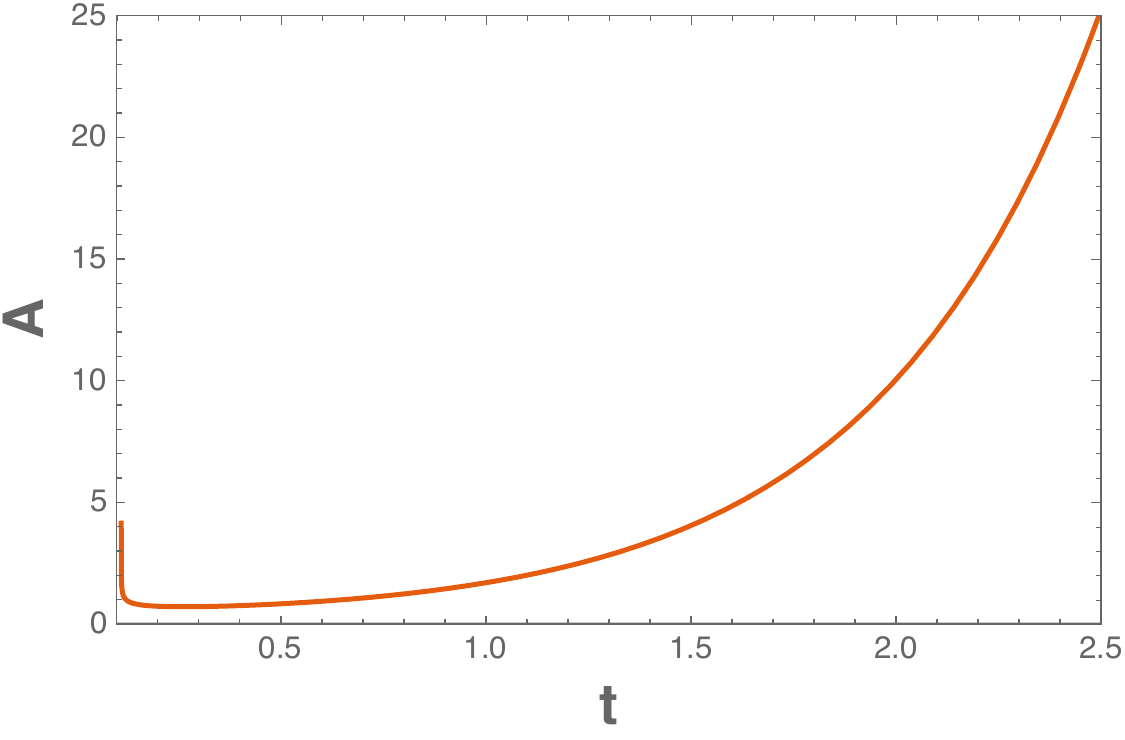}
\includegraphics[scale=0.3]{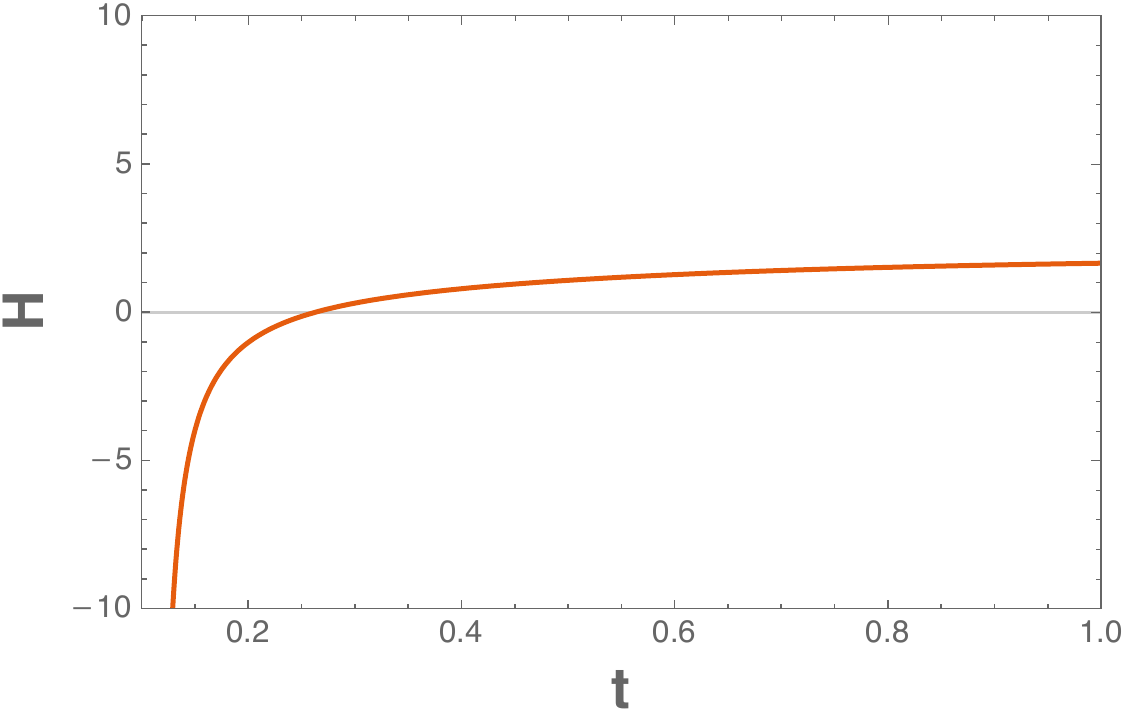}
\includegraphics[scale=0.3]{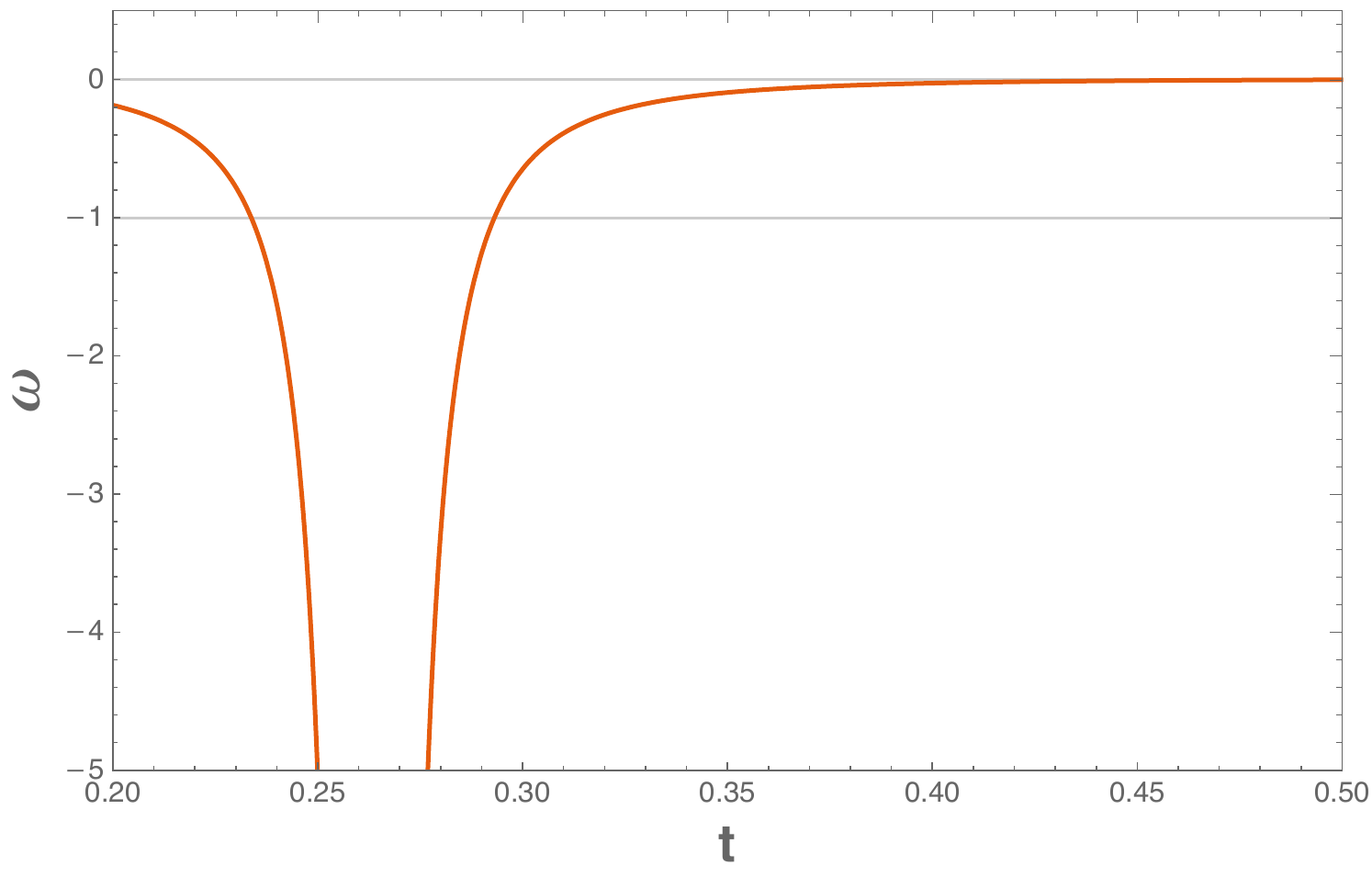}
\caption{ Phantom domination. This figure shows the time ($ 0.1\leq
t\leq 2.5$) evolution of the scale factor $A(t)$, the~Hubble
parameter $H(t)$, and~the barotropic parameter $\omega_{\phi_1
\phi_2}(t)$. We use arbitrary units, namely, $V_1=V_2=1.0$,
$r_1=1.5$, $r_2=0.9$, $q_1=q_2=0.1$, $a_3=-6.0$,
$\lambda_{2}=\sqrt{2}$, and~$\lambda_{1} = 6/\lambda_{2}$.
The~remaining constants can be obtained from the aforementioned
values. Note that time is measured in reduced Planck units since
$8\pi G=1$. } \label{figura_2}
\end{figure}

In Figure~\ref{figura_2}, we can appreciate the evolution of the
scale factor, the Hubble parameter, and~the barotropic parameter,
with~respect to time. First, we can once again observe a bouncing
$A$, which consolidates our previous outcome. In~fact, this behavior
was claimed recently in \cite{genly2022}, using a dynamical system
approach. Additionally, in~the upper right plot, $H$ crosses the
horizontal axis (at the bounce of $A$). Then, in the panel at the
bottom, once again $\omega_{\phi_1 \phi_2}$ traverses the phantom
divide line ``$-$1'', an~upshot consistent with the quintom
description.

%
\subsubsection{Quintessence Domination: $\lambda_1<\sqrt{6}$
and $\lambda_2>\sqrt{6}$.}
We reinsert the solutions of $\lambda_1<\sqrt{6}$ ($\eta_1>0$) and
$\lambda_2>\sqrt{6}$ into the Hamilton equations for the momenta,
leading to
\begin{eqnarray}
P_1 &=& p_1 - \frac{r_1}{12\eta_1} \, Coth\left(r_1 t-q_1
\right) \,, \label{solution-p1-1} \\
P_2 &=& p_2 - \frac{r_2}{12\eta_2} \, Coth\left(r_2 t-q_2 \right)
\,, \label{solution-p2-1}
\end{eqnarray}
where $p_1$ and $p_2$ are integration constants. We use
(\ref{solution-p1-1}) and (\ref{solution-p2-1}) to obtain a null
Hamiltonian when
\begin{equation}
p_1=\frac{\eta_1-9}{\eta_1} p_3, \qquad p_2=\frac{\eta_2-9}{\eta_2}
p_3,\qquad p_3= \pm\frac{1}{36} \sqrt{\frac{\eta_2 r_1^2+\eta_1
r_2^2 }{3\left[3\eta_1 +3\eta_2 -\eta_1\eta_2\right]}} \,.
\label{coefficients1}
\end{equation}

As a consequence, the~solutions of $\xi_i$ take the following form:
\begin{align}
\xi_1 &= Ln\left(\frac{r_1^2}{288\eta_1\, V_1}\right)+Ln\left[
Csch^2
\left(r_1\,t -q_1 \right)\right] \,,  \label{xi1-11}\\
\xi_2 &= Ln\left(\frac{r_2^2}{288\eta_2\, V_2}\right)+Ln\left[
Csch^2
\left(r_2\,t -q_2 \right)\right] \,, \label{xi2-11}\\
\xi_3 &= a_3 - 648\frac{-\eta_1\, \eta_2+3\eta_1 +
3\eta_2}{\eta_1\,\eta_2} p_3 t
+\frac{\eta_1-9}{\eta_1}Ln\left[Sinh^2\left(r_1\,t - q_1 \right)
\right]\nonumber\\
& \quad +\frac{\eta_2-9}{\eta_2}Ln\left[Sinh^2\left(r_2\,t - q_2
\right) \right] \,,
\end{align}
with an integration constant $a_3$. Then, we apply the inverse
transformation (\ref{trans_2}) to arrive at the solutions in terms
of the original variables, which read
\begin{equation}\label{sols_quintessence}
\begin{split}
\Omega &=  Ln\left[\frac{r_1 r_2}{288\sqrt{\eta_1 \eta_2 V_1 V_2}}
\right]^{\frac{1}{9}}+\frac{a_3}{18} + Ln\left[ Csch^{\beta_1}
\left(r_1\,t -q_1 \right) Csch^{\beta_2}\left(r_2\, t -q_2
\right)\right] \nonumber\\
& \quad -36\frac{3\eta_1 + 3\eta_2-\eta_1\,
\eta_2}{\eta_1\,\eta_2} p_3 t \,,\\
 \phi_1 &=  \phi_{10}+ Ln\left[Cosh^{\frac{2\left(|\eta_1|+3 \right)}{\lambda_1 |\eta_1|}}\left(r_1\,t-q_1\right)
Csch^{\frac{6}{\lambda_1 \eta_2}}\left(r_2\, t
-q_2\right)\right]+216\frac{|\eta_1|\, \eta_2-3|\eta_1| +
3\eta_2}{\lambda_1\, |\eta_1|\,\eta_2} p_3 t  \,, \\
 \phi_2 &=\phi_{20}+Ln\left[Cosh^{\frac{6}{\lambda_2 |\eta_1|}}\left(r_1\,t-q_1\right) Sinh^{\frac{2\left(
\eta_2-3\right)}{\lambda_2 \eta_2}}\left(r_2\, t
-q_2\right)\right]\nonumber\\
& \quad +216\frac{|\eta_1|\, \eta_2-3|\eta_1| + 3\eta_2}{\lambda_2\,
|\eta_1|\,\eta_2} p_3 t  \,,
\end{split}
\end{equation}
where $\beta_1=1/\eta_1$, $\beta_2=1/\eta_2$, and~the constants
$\Omega_0, \phi_{10}$, and~$\phi_{20}$ are those in (\ref{const}).
With~these solutions, we can write the scale factor in the following
form:
\begin{align}
A(t)&= \left[\frac{r_1 r_2}{288\sqrt{\eta_1\eta_2 V_1 V_2}}
\right]^{\frac{1}{9}}\,e^{\frac{a_3}{18}}\, Csch^{\beta_1}
\left(r_1\,t -q_1 \right) Csch^{\beta_2}\left(r_2\, t -q_2
\right)\,\nonumber\\
& \quad \times  Exp\left[-36\frac{3\eta_1 + 3\eta_2-\eta_1\,
\eta_2}{\eta_1\,\eta_2} p_3 t\right] \,.\label{scale-factor}
\end{align}

Immediately, one can observe that to obtain an increasing scale
factor with respect to time, the~constant $p_3$ must be negative.
However, none of the parameters considered in this scenario lead to
$p_3 <0$; therefore, this solution is not physically relevant.
%
%

\section{Quantum~Formalism}\label{section3}
To present the quantum mechanical version of the classical model,
in~(\ref{hamifrw}) we promote the classical momenta to operators
making the replacement $\Pi_{q^\mu}=-i\hbar\partial_{q^\mu}$,
obtaining the following Hamiltonian density: \vspace{-6pt}

\begin{equation}
{\cal H}=\Pi_\Omega^2 +Qi\hbar \Pi_\Omega-12\Lambda_2
\Pi_{\phi_1}^2+12\Lambda_1 \Pi_{\phi_2}^2 -24 \Lambda_0
\Pi_{\phi_1}\Pi_{\phi_2} -24V_1 e^{-\lambda_1\phi_1+6\Omega} -24V_2
e^{-\lambda_2\phi_2+6\Omega}. \label{q-hamifrw-mod}
\end{equation}

To obtain Equation~(\ref{q-hamifrw-mod}), we have substituted
$e^{-3\Omega}\Pi_\Omega^2 \to e^{-3\Omega}\left[\Pi_\Omega^2+Qi\hbar
\Pi_\Omega \right]$ since one has to take into account the
factor-ordering problem between the $e^{-3\Omega}$ and its momentum
$\Pi_\Omega$; hence, $Q$ is a number that measures such ambiguity.
In~order to have a more manageable functional form of
(\ref{q-hamifrw-mod}), we take the constraint of the matrix element
$m^{12}$ (Equation (\ref{constraint})); then, we apply the canonical
transformation on variables $(\Omega,\phi_1,\phi_2)\leftrightarrow
(\xi_1,\xi_2,\xi_3)$ (Equations (\ref{trans_2}) and
(\ref{new-moment})), as~well as the gauge $N=24e^{3\Omega}$.
Therefore, we obtain
\begin{align}
{\cal H} &= 12 \eta_1 P_1^2+12\eta_2P_2^2+ 12\left(-9 +
\eta_1+\eta_2\right)P_3^2  \nonumber\\
& \quad + 24P_3\left[ \left(9-\eta_1 \right)P_1+ \left(9
-\eta_2\right)P_2\right]
  +6Qi\hbar \left(P_1+P_2+P_3\right)\quad - 24\left(V_1 e^{\xi_1}+V_2
e^{\xi_2}\right),\label{q-hamiltonian}
\end{align}
with $\eta_1=3-\Lambda_1$ and $\eta_2=3+\Lambda_2$. Recall that the
Hamiltonian density is identically zero ${\cal H}=0$; hence, the
quantum counterpart of (\ref{q-hamiltonian}) is obtained by applying
the same prescription used to obtain (\ref{q-hamifrw-mod}). Having
this at hand, we can write down the Wheeler--DeWitt (WDW) equation,
which reads \vspace{-6pt}
\begin{align}
\hat{\cal H}\Psi(\xi_i) &=  -12\hbar^2 \eta_1 \frac{\partial^2
\Psi}{\partial \xi_1^2}-12\hbar^2 \eta_2\frac{\partial^2
\Psi}{\partial \xi_2^2}- 12\hbar^2\left(-9 +
\eta_1+\eta_2\right)\frac{\partial^2 \Psi}{\partial
\xi_3^2}\nonumber\\
& \quad +6Q\hbar^2 \left(\frac{\partial \Psi}{\partial \xi_1}
+\frac{\partial \Psi}{\partial \xi_2}+\frac{\partial \Psi}{\partial
\xi_3}\right)
 -\hbar^2 24\left[ \left(9-\eta_1 \right)\frac{\partial^2
\Psi}{\partial \xi_1 \partial \xi_3}+ \left(9
-\eta_2\right)\frac{\partial^2 \Psi}{\partial \xi_2 \partial
\xi_3}\right]
\nonumber\\
&\quad   - 24\left(V_1 e^{\xi_1}+V_2 e^{\xi_2}\right)\Psi=0
\,.\label{wdw}
\end{align}

In order to solve the WDW equation, we propose the following
solution for the wave function $ \Psi(\xi_1,\xi_2,\xi_3)=e^{p_3
\xi_3}{\cal G}(\xi_1,\xi_2)$ with $ p_3=constant$. Additionally, we
take as an ansatz ${\cal G}(\xi_1,\xi_2)= G_1(\xi_1) G_2(\xi_2)$;
upon substitution in (\ref{wdw}), we obtain the following.
\begin{align}
 &-12 \eta_1 G_2\frac{\partial^2 G_1}{\partial
\xi_1^2}+6\left(Q -4p_3(9-\eta_1)\right)G_2\frac{\partial
G_1}{\partial
\xi_1}\nonumber\\
& \quad+3\,\left[p_3\left(Q-2p_3(9-\eta_1+\eta_2)
\right)-8\frac{V_1}{\hbar^2}
e^{\xi_1}\right]G_1G_2 +\nonumber\\
&\quad -12 \eta_2 G_1\frac{\partial^2 G_2}{\partial
\xi_2^2}+6\left(Q -4p_3(9-\eta_2)\right)G_1\frac{\partial
G_2}{\partial
\xi_2}\nonumber\\
&\quad +3\,\left[p_3\left(Q-2p_3(9-\eta_1+\eta_2)
\right)-8\frac{V_2}{\hbar^2} e^{\xi_2}\right]G_1G_2=0
\,,\label{quantum}
\end{align}
finally, we factorize $G_1 G_2$. Thus, two ordinary differential
equations for the functions $G_1$ and $G_2$ emerge
\begin{multline}
 -12\frac{ \eta_1}{G_1}\frac{\partial^2 G_1}{\partial
\xi_1^2}+6\left(Q -4p_3(9-\eta_1)\right)\frac{1}{G_1}\frac{\partial
G_1}{\partial\xi_1} \\
+3\left[p_3\left(Q-2p_3(9-\eta_1+\eta_2)\right)-8\frac{V_1}{\hbar^2}e^{\xi_1}\right]
-\nu^2= 0 ,\label{quantum-1}
\end{multline}
\begin{multline}
 -12\frac{ \eta_2}{G_2}\frac{\partial^2 G_2}{\partial
\xi_2^2}+6\left[Q-4p_3(9-\eta_2)\right]\frac{1}{G_2} \frac{\partial
G_2}{\partial \xi_2}\\
 \quad  +3\left[p_3\left(Q-2p_3(9-\eta_1+\eta_2)
\right)-8\frac{V_2}{\hbar^2} e^{\xi_2}\right]+ \nu^2 =
0,\label{quantum-2}
\end{multline}
where $\nu^2$ is an arbitrary constant. These last two equations can
be written as $y^{\prime \prime} + a y^\prime + \left(b e^{\kappa x
} +c \right)y=0$, and~their solutions are of the form
~\cite{polyanin}
\begin{equation}
Y(x)=Exp\left({-\frac{ax}{2}}\right)
Z_\rho\left(\frac{2\sqrt{b}}{\kappa} e^{\frac{\kappa x}{2}}\right)
\,,
\end{equation}
here, $Z_\rho$ are the generic Bessel functions with the order $
\rho=\sqrt{a^2-4c}/\kappa$. If~$\sqrt{b}$ is real, $Z_\rho$ becomes
the ordinary Bessel function; otherwise, the~solutions will be given
in terms of the modified Bessel functions. In~the next sections, we
will show quantum solutions separated into two classes, according to
$\eta_1$ and $\lambda_1\lambda_2=6$.

%
%
\subsection{Quantum Solution for $\eta_1 >0$ and $\lambda_1<\sqrt{6}$}
First, we identify the following expressions for
Equation~(\ref{quantum-1}):
\begin{eqnarray}
\kappa &=& 1 \,, \qquad
 a =  -\frac{Q -4p_3(9-\eta_1)}{2\eta_1} \,,\nonumber\\
b &=&  \frac{2V_2}{\eta_1 \hbar^2} \,, \qquad c =
-\frac{p_3\left[Q-2p_3(9-\eta_1+\eta_2)\right]}{4\eta_1}+\frac{\nu^2}{12\eta_1}
\,, \label{idem11}
\end{eqnarray}
and for (\ref{quantum-2})
\begin{eqnarray}
 \kappa &=& 1 \,, \qquad
a =  -\frac{Q -4p_3(9-\eta_2)}{2\eta_2} \,,\nonumber\\
b &=& \frac{2V_1}{\eta_2 \hbar^2} \,, \qquad  c =
-\frac{p_3\left[Q-2p_3(9-\eta_1+\eta_2)\right]}{4\eta_2}-\frac{\nu^2}{12\eta_2}
\,. \label{idem12}
\end{eqnarray}

Note that in both cases, $\sqrt{b}$ is real; then, the solutions are
written in terms of the ordinary Bessel functions $
Z_{\rho_i}=J_{\rho_i}$. Thus, the~wave function becomes the
following:
\begin{equation}
{\cal B}_{\rho_1 \rho_2} = {\cal B}_0\, J_{\rho_1} \left[
\frac{2}{\hbar}\sqrt{\frac{2V_1}{\eta_1}} e^{\frac{\xi_1}{2}}
\right] J_{\rho_2}\left[\frac{2}{\hbar}\sqrt{\frac{2V_2}{\eta_2}}
e^{\frac{\xi_2}{2}}
 \right]\,\,e^\theta \,, \nonumber
\end{equation}
where $$\theta=\frac{Q -4p_3(9-\eta_1)}{4\eta_1}\xi_1 + \frac{Q
-4p_3(9-\eta_2)}{4\eta_2}\xi_2 \,,\nonumber$$ and ${\cal B}_0$ is an
integration constant. Additionally, the~order of the two Bessel
functions are
\begin{eqnarray}
&&\rho_1=\sqrt{\left(-\frac{Q-4p_3(9-\eta_1)}{2\eta_1}
\right)^2+\frac{p_3(Q-2p_3(9-\eta_1+\eta_2))}{\eta_1}-\frac{\nu^2}{3\eta_1}} \,,\\
&& \rho_2=\sqrt{\left(-\frac{Q-4p_3(9-\eta_2)}{2\eta_2}
\right)^2+\frac{p_3(Q-2p_3(9-\eta_1+\eta_2))}{\eta_2}+\frac{\nu^2}{3\eta_2}}
\,.
\end{eqnarray}

Hence, the~wave function $\Psi$ in the original variables becomes
\begin{equation}\small
\Psi_{\rho_1 \rho_2} =
\Psi_0\,A^{6\alpha}Exp\left[\alpha_1\,\lambda_1 \phi_1
+\alpha_2\,\lambda_2 \phi_2\right] J_{\rho_1} \left[
\frac{2}{\hbar}\sqrt{\frac{2V_1}{\eta_1}} A^3\, e^{\frac{-\lambda_1
\phi_1}{2}} \right]
J_{\rho_2}\left[\frac{2}{\hbar}\sqrt{\frac{2V_2}{\eta_2}}
A^3e^{-\frac{\lambda_2 \phi_2}{2}}
 \right] \,,
\label{wa}
\end{equation}
where $\Psi_0$ is a normalization constant, and~%
\begin{eqnarray}
&& \alpha=\frac{Q(\eta_2+\eta_1)}{4\eta_1\eta_2}-\frac{p_3(9-\eta_1)}{\eta_1}-\frac{p_3(9-\eta_2)}{\eta_2}+p_3 \,, \nonumber\\
&& \alpha_1= -\frac{Q -4p_3(9-\eta_1)}{4\eta_1}+p_3 \,,\quad
\alpha_2=- \frac{Q -4p_3(9-\eta_2)}{4\eta_2}+p_3 \,.
\end{eqnarray}

By analyzing solution (\ref{wa}), we could not find any set of
parameter values for which the probability density function (defined
by the wave function (\ref{wa})) is bounded. This unwanted behavior
prevents us from directly implementing the standard interpretation
of quantum mechanics in order to draw meaningful physical
conclusions. This setback is tempered by the fact that the
corresponding classical solution (given essentially by
(\ref{scale-factor})) is not of physical relevance, and~so no
further analysis  will be performed regarding this~case.


\subsection{Quantum solution when $\eta_1 <0$ and $\lambda_1 > \sqrt{6}$.}
We set up the corresponding parameters for
Equation~(\ref{quantum-1})
\begin{eqnarray}
\kappa &=& 1 \qquad a = \frac{Q -4p_3(9+|\eta_1|)}{2|\eta_1|},\nonumber\\
b &=&  -\frac{2V_2}{|\eta_1| \hbar^2} \qquad
 c =\frac{p_3\left[Q-2p_3(9+|\eta_1|+\eta_2)\right]}{4|\eta_1|}-\frac{\nu^2}{12|\eta_1|} \,,
\label{idem21}
\end{eqnarray}
and for (\ref{quantum-2})
\begin{eqnarray}
 \kappa &=& 1 \qquad
 a =  -\frac{Q -4p_3(9-\eta_2)}{2\eta_2} \,,\nonumber\\
 b &=& \frac{2V_1}{\eta_2 \hbar^2},\qquad
 c =-\frac{p_3\left[Q-2p_3(9+|\eta_1|+\eta_2)\right]}{4\eta_2}-\frac{\nu^2}{12\eta_2} \,,
\label{idem22}
\end{eqnarray}
note that we have inverted the sign of the previous formulas.
Hereby, we introduce $ |\eta_1|$. Then, the~first case
(\ref{idem21}) yields an imaginary $\sqrt{b}$; therefore, its
solution must be in terms of the modified Bessel function
$Z_{\rho_1}=K_{\rho_1}$ (contrary to the second case, where the
proper function is $Z_{\rho_2}=J_{\rho_2}$). Hence, we have
\begin{equation}
 {\cal B}_{\rho_1 \rho_2} ={\cal B}_0\, K_{\rho_1} \left[
\frac{2}{\hbar}\sqrt{\frac{2V_1}{|\eta_1|}} e^{\frac{\xi_1}{2}}
\right] J_{\rho_2}\left[\frac{2}{\hbar}\sqrt{\frac{2V_2}{\eta_2}}
e^{\frac{\xi_2}{2}}
 \right]\,\, e^{\theta_2},
\end{equation}
here
\begin{equation}
\theta_2=-\frac{Q -4p_3(9+|\eta_1|)}{4|\eta_1|}\xi_1 + \frac{Q
-4p_3(9-\eta_2)}{4\eta_2}\xi_2 \,,
\end{equation}
and the order of both Bessel functions are
\begin{eqnarray}
&& \rho_1=\sqrt{\left(\frac{Q-4p_3(9+|\eta_1|}{2|\eta_1|}
\right)^2-\frac{p_3(Q-2p_3(9+|\eta_1|+\eta_2))}{|\eta_1|}+\frac{\nu^2}{3|\eta_1|}} \,, \\
&& \rho_2=\sqrt{\left(-\frac{Q-4p_3(9-\eta_2}{2\eta_2}
\right)^2+\frac{p_3(Q-2p_3(9+|\eta_1|+\eta_2))}{\eta_2}+\frac{\nu^2}{3\eta_2}}
\,.
\end{eqnarray}

\begin{figure}[ht!]
\begin{center}
\includegraphics[scale=0.35]{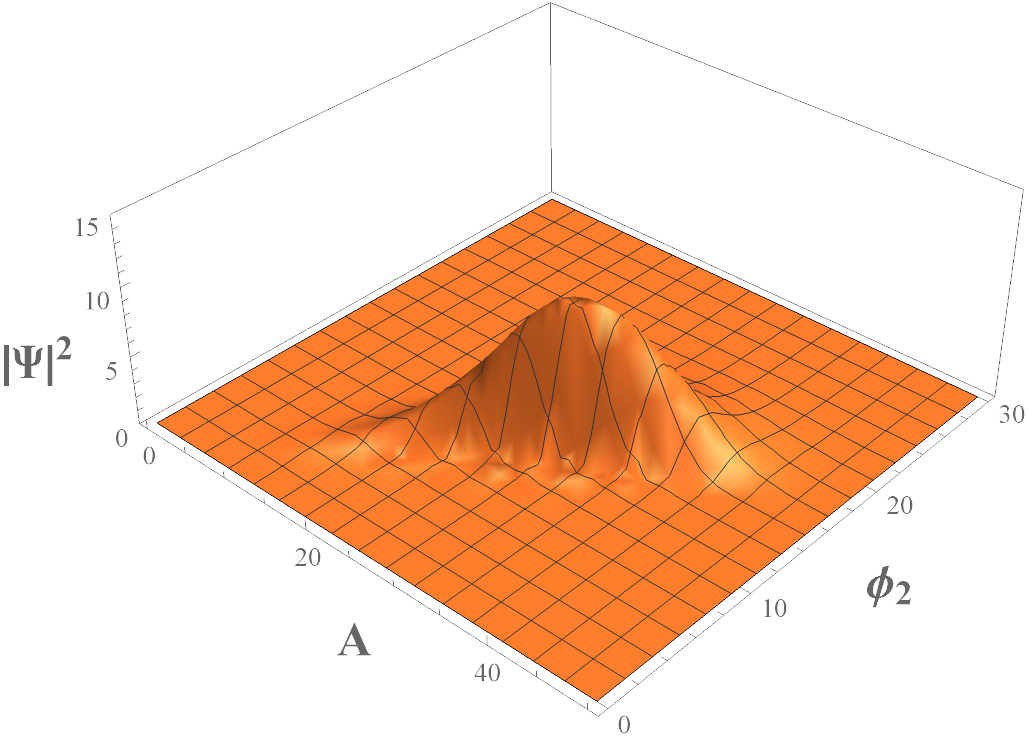}
\includegraphics[scale=0.35]{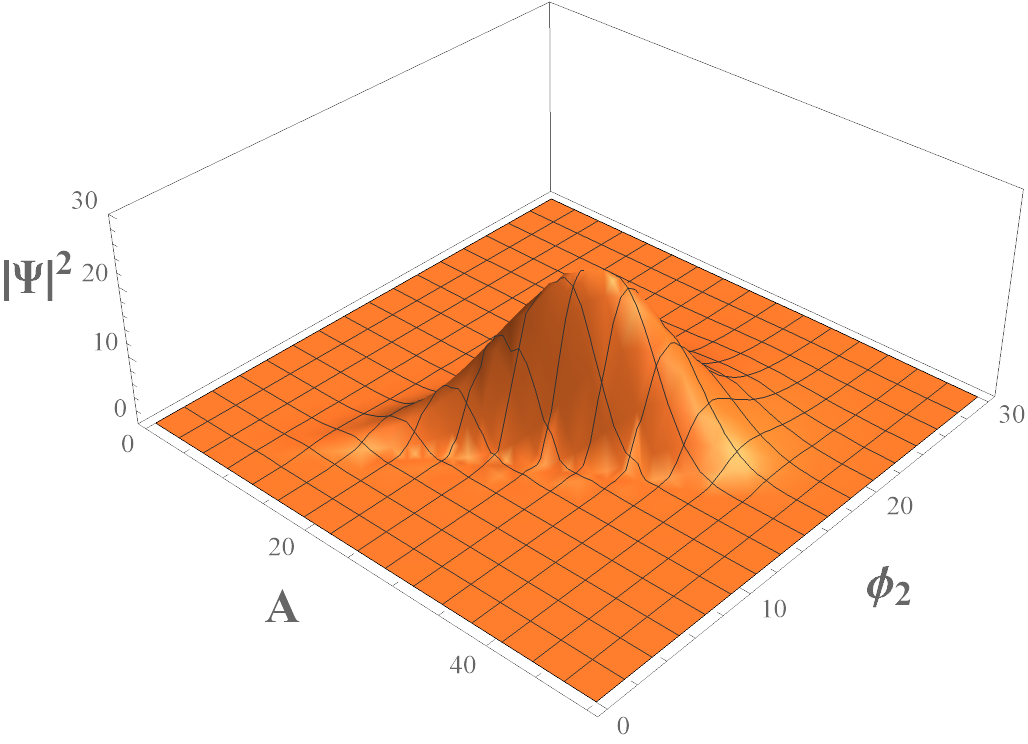}
\includegraphics[scale=0.35]{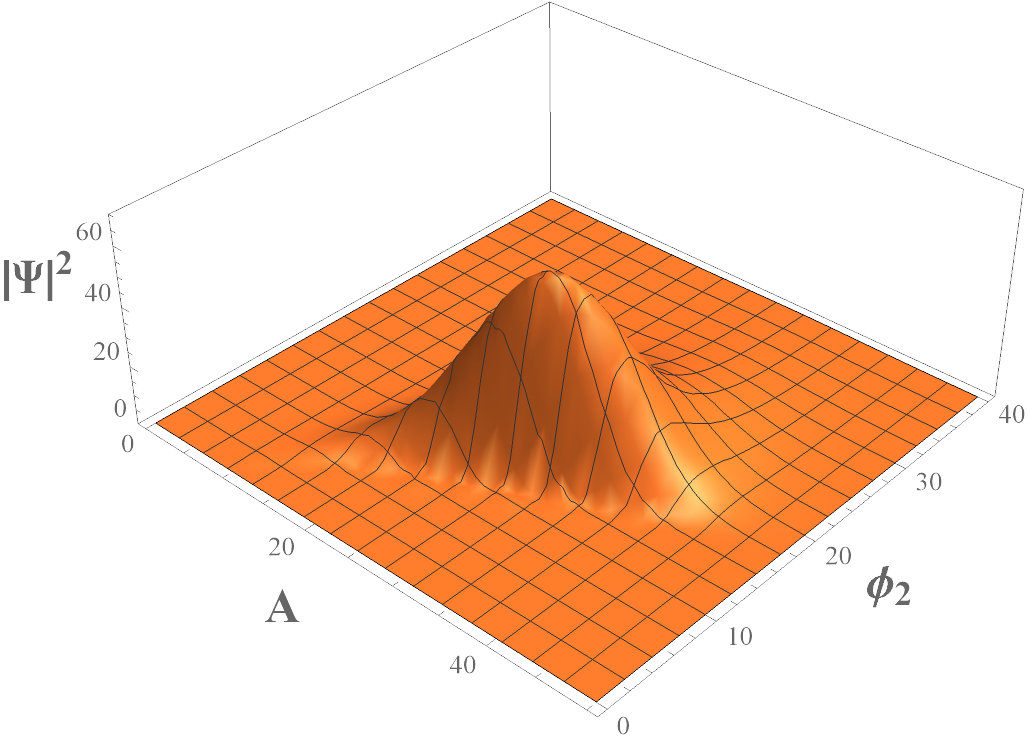}
\includegraphics[scale=0.35]{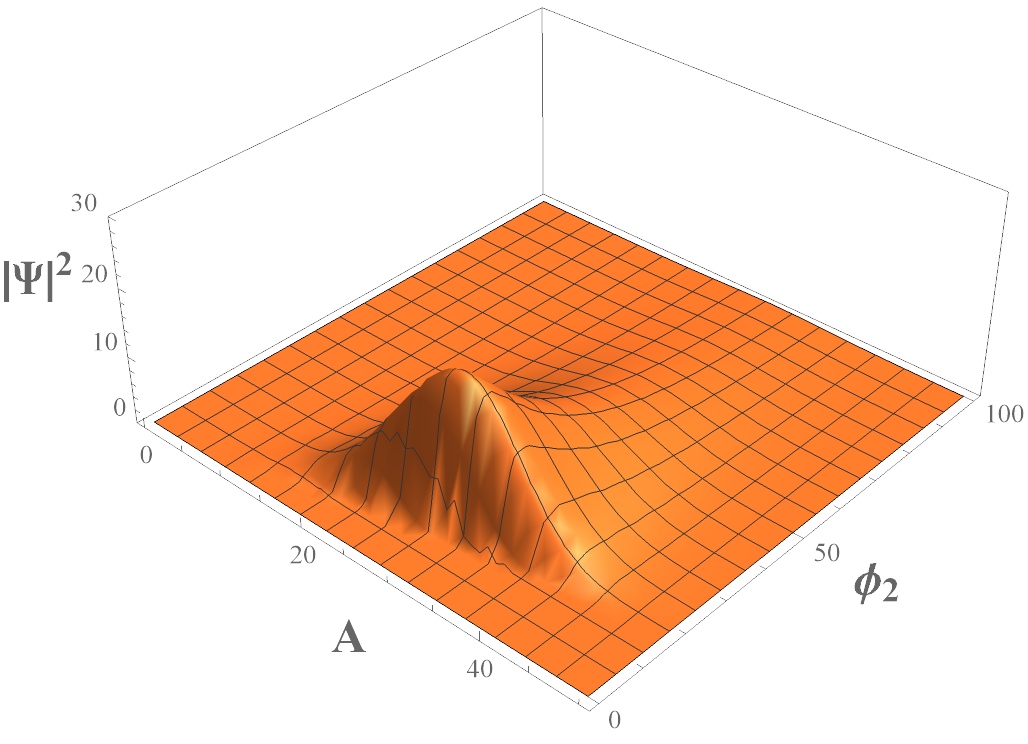}
\caption{Phantom scenario. These figures show the probability
density of the wave function (\ref{wavefunction}) for the values of
$Q=2$ and $Q=0$ (top panels from left to right, respectivley), and~$
Q=-2$  and $ Q=-4$ (bottom panels from left to right, respectively).
We use arbitrary units, namely, $\nu=10$, $\lambda_1=10.5$, $
\lambda_2=6/\lambda_1$, $ V_1 = 0.1 \,, V_2 = 10^{-5}$, $ r_1=49$, $
r_2= 1.5 \,, a_3=-2.3$, and $ p_{3}=0.326878$, and~the bounce in the
quintessence field $\phi_1=1.455$. The~remaining constants can be
obtained from the aforementioned values. Additionally, for~$ Q=2,0$
we take $\Psi_0=10^{-3}\,, 10^{-2}$ respectively; then, for $
Q=-2\,,-4$ we chose $\Psi_0=10^{-1},1/\sqrt{10}$ respectively. Note
that the probability density tends toward the phantom sector when
the factor ordering constant $ Q\ll 0$.} \label{figura_variosQ}
\end{center}
\end{figure}

Finally, the~wave function in the original variables is given by
\begin{equation}
\Psi_{\rho_1 \rho_2} = \Psi_0 A^{6\beta}Exp\left[\alpha_1\,\lambda_1
\phi_1+\alpha_2\,\lambda_2 \phi_2\right] K_{\rho_1} \left[
\frac{2}{\hbar}\sqrt{\frac{2V_1}{|\eta_1|}} A^3\,
e^{\frac{-\lambda_1 \phi_1}{2}} \right]
J_{\rho_2}\left[\frac{2}{\hbar}\sqrt{\frac{2V_2}{\eta_2}}
A^3e^{-\frac{\lambda_2 \phi_2}{2}} \right] \,, \label{wavefunction}
\end{equation}
where
\begin{eqnarray}
&&
\beta=-\frac{Q(\eta_2-|\eta_1|)}{4|\eta_1|\eta_2}+\frac{p_3(9+|\eta_1|)}{|\eta_1|}-\frac{p_3(9-\eta_2)}{\eta_2}+p_3 \,, \\
&& \alpha_1=\frac{Q -4p_3(9+|\eta_1|)}{4|\eta_1|}+p_3  \,, \quad
\alpha_2= - \frac{Q -4p_3(9-\eta_2)}{2\eta_2}+p_3 \,,
\end{eqnarray}
and a normalization constant $\Psi_0$. The~behaviour of the
probability density can be seen in Figure~\ref{figura_variosQ}.
 Observe that in all panels, the probability density dies away as the scale factor and scalar field evolve,
 an~expected outcome already reported in~\cite{Socorro:2020nsm, Socorro:2019wpu, Socorro:2018amv}.
 On~the other hand, we vary the factor ordering constant $Q$, in~order to show how $ |\Psi|^{2}$
 behaves. We can see that whilst $Q\ll 0$, the probability density tends to the phantom sector.
 In~\cite{Socorro:2020nsm}, the authors showed that the parameter $Q$ acts a retarder of the wave
 function and compresses the length on the axis where the field evolves; however, they analysed the
 case of two  quintessence~fields.

%
%
\subsection{Quantum Solution when $\lambda_1 =\lambda_2=\sqrt{6}$,
Therefore $\eta_1=0$ and $\eta_2=6$.}
In this final case, we take $ \lambda_1=\lambda_2=\sqrt{6}$;
therefore, $\eta_1=0$ and $ \eta_2=6$. Hence, the
Equations~(\ref{quantum-1}) and (\ref{quantum-2}) can be reduced to
\begin{align}
 6\left(Q -36p_3\right)\frac{1}{G_1}\frac{\partial G_1}{\partial
\xi_1}+3\left[p_3\left(Q-30p_3 \right)-8\frac{V_1}{\hbar^2}
e^{\xi_1}\right] -\nu^2 &= 0 \,,\label{quantum-11}\\
-\frac{72 }{G_2}\frac{\partial^2 G_2}{\partial
\xi_2^2}+6\left[Q-12p_3\right]\frac{1}{G_2} \frac{\partial
G_2}{\partial \xi_2} +3\left[p_3\left(Q-30p_3
\right)-8\frac{V_2}{\hbar^2} e^{\xi_2}\right]+ \nu^2 &=\rm 0 \,.
\label{quantum-22}
\end{align}

The solution of (\ref{quantum-11}) is given by
\begin{equation}
 G_1=G_0
Exp\left[\frac{\frac{\nu^2}{3}-p_3(Q-30p_3)}{2(Q-36p_3)}\xi_1 +
\frac{4V_1}{\hbar^2(Q-36p_3)} e^{\xi_1} \right] \,, \label{g1}
\end{equation}
where $G_0$ in an integration constant. Then, for~$G_2$ we have the
following ordinary Bessel function:
\begin{equation}
 G_2=Exp\left[\frac{Q-12p_3}{24}\xi_2 \right]\,
J_{\rho_2}\left[\frac{\sqrt{V_2}}{\hbar}e^{\frac{\xi_2}{2}} \right],
\label{g2}
\end{equation}
here, the order is
\begin{equation}
\rho_2= \sqrt{\left(\frac{Q-12p_3}{12}
\right)^2+\frac{1}{6}\left[\frac{\nu^2}{3}+p_3(Q-30p_3) \right]} \,.
\end{equation}
Remarkably for this case, we can obtain a parameter space of $Q\,,
\nu$, and~$p_3$ where the order can be real or imaginary. Hence, we
have
\begin{equation}
{\cal B}= {\cal B}_0 \,
J_{\rho_2}\left[\frac{\sqrt{V_2}}{\hbar}e^{\frac{\xi_2}{2}}
\right]\,\, e^{\theta_3},
\end{equation}
with \begin{equation}
\theta_3=\frac{\frac{\nu^2}{3}-p_3(Q-30p_3)}{2(Q-36p_3)}\xi_1
+\frac{Q-12p_3}{24}\xi_2+ \frac{4V_1}{\hbar^2(Q-36p_3)} e^{\xi_1}
\,,\nonumber
\end{equation}

Finally, in~the original variables the wave function is
\begin{equation}
\Psi=\psi_0
A^{6\eta}J_{\rho_2}\left[\frac{\sqrt{V_2}}{\hbar}A^3e^{\frac{-\lambda_2
\phi_2}{2}} \right]\,\, e^{\theta_1} \,, \label{lastWF}
\end{equation}
where
\begin{equation}
\theta_1=\frac{4V_1}{\hbar^2(Q-36p_3)}\,A^6 e^{-\lambda_1 \phi_1}+
\alpha_1 \,\lambda_1 \phi_1 +\alpha_2\,\lambda_2 \phi_2 \,,
\end{equation}
and
\begin{eqnarray}
&& \hspace{2cm}
\eta=\frac{\frac{\nu^2}{3}-p_3(Q-30p_3)}{2(Q-36p_3)}+\frac{Q-12p_3}{24}+p_3 \nonumber\\
&&\alpha_1= -\frac{\frac{\nu^2}{3}-p_3(Q-30p_3)}{2(Q-36p_3)}+p_3
\,,\quad \alpha_2=-\frac{Q-12p_3}{24} +p_3 \,,
\end{eqnarray}
and a normalization constant $\psi_0$. For~completeness of the above
classical solutions, we include this case; however, once more the
probability density function is not bounded since $ |\Psi|^{2}$ does
not fade as the scale factor and scalar field evolve. We recall that
the standard interpretation of quantum mechanics becomes troublesome
to realize due to this nuisance behavior.
 Therefore, the~wave function (\ref{lastWF}) is not physically relevant.
%
%
\section{Final~Remarks}\label{section4}
In this work, we have studied a chiral cosmological model from the
point of view of a K-essence formalism. The background geometry was
a flat FLRW universe minimally coupled to quintom fields: one
quintessence and one phantom. In~this approach, the~scalar fields
interact within the kinetic and potential~sectors.

In the classical framework, we established the Hamiltonian density
(\ref{hamiltonian}), which in turn allows
 one to find exact solutions for different sets of values of the free parameters. We highlight two cases: the~ first
  when $\lambda_1=\lambda_2=\sqrt{6}$, and the second where phantom domination is the relevant factor,
 namely, $ \lambda_1>\sqrt{6}$ and $\lambda_2<\sqrt{6}$. In~the two scenarios, the~scale factor grows very rapidly
 and the big-bang singularity is avoided via a bounce. We call it the ``big bounce''. In~fact, this claim is also
 supported by the behavior of both the scale factor and the Hubble parameter. Finally, we show that the barotropic
 parameter is capable of transiting from a quintessence phase to a phantom one, i.e.,~it crosses the phantom divide line.
 In~Figures~\ref{figura_1} and \ref{figura_2}, we show the behavior of these quantities as a function of~time.

On the other hand, using the canonical quantization procedure, we
were able to establish the quantum
 counterpart of the classical model and~compute the Wheeler--DeWitt equation. Once again, we solve it
  for various scenarios given by different sets of values of the free parameters. In~particular, we
  found exact solutions for three distinct cases: $\eta_1 > 0$ and $\lambda_1 < \sqrt{6}$, $ \eta_1 < 0$
  and $\lambda_1 > \sqrt{6}$, and $ \lambda_1=\lambda_2=\sqrt{6}$; therefore, $\eta_1=0$ and $\eta_2=6$.
  Figure~\ref{figura_variosQ} shows the behavior of the probability density as a function of the scale
   factor and scalar field, for~the phantom case, i.e.,~$ \eta_1 < 0$ and $ \lambda_1 > \sqrt{6}$.
   The probability density exhibits a damped behavior as the scale factor and scalar fields evolve.
   An expected result has already been reported in~\cite{Socorro:2020nsm, Socorro:2019wpu, Socorro:2018amv}.
   Lastly, we note that by varying $Q$, specifically when $Q\ll 0$, the~probability density evolves towards
   the phantom sector. This outcome contrasts with that reported in~\cite{Socorro:2020nsm}, where the authors
   showed that the parameter $Q$ delays the evolution of the wave function and compresses the length on the
   axis where the field evolves; however, they analyzed the case of two quintessence~fields.

\vspace{6pt} \noindent {\bf author contributions}{{: {\it
Conceptualization}, J. Socorro, Sinuhe P\'erez Pay\'an, Rafael
Hern\'andez-Jim\'enez, Abraham Espinoza Garc\'ia, and Luis Rey D\'az
Barr\'on; {\it Methodology}, J. Socorro, Sinuhe P\'erez Pay\'an,
Rafael Hern\'andez-Jim\'enez, Abraham Espinoza Garc\'ia, and Luis
Rey D\'iaz Barr\'on; {\it Writing - Original Draft}, J. Socorro,
Sinuhe P\'erez Pay\'an, Rafael Hern\'andez-Jim\'enez, Abraham
Espinoza Garc\'ia, and Luis Rey D\'iaz Barr\'on; {\it Writing -
Review and Editing}, J. Socorro, Sinuhe P\'erez Pay\'an, Rafael
Hern\'andez-Jim\'enez, Abraham Espinoza Garc\'ia, and Luis Rey
D\'iaz Barr\'on; {\it Visualization}, J. Socorro, Sinuhe P\'erez
Pay\'an. All authors have read and agreed to the published version
of the manuscript. }}
\bigskip

\noindent {\bf funding:}{ This work was partially supported by
PROMEP grants UGTO-CA-3. J.S. and L. R. D. B. were partially
supported SNI-CONACyT. R.H.J is supported by CONACyT Estancias
posdoctorales por M\'exico, Modalidad 1: Estancia Posdoctoral
Acad\'emica.}

\bigskip
\noindent {\bf data availability:}{{ Not applicable }.}

\bigskip
\noindent {\bf conflicts of interest:}{{The authors declare no
conflict of interest}.}

\bigskip
\acknowledgments{ This work is part of the collaboration within the
Instituto Avanzado de Cosmolog\'{\i}a and Red PROMEP: Gravitation
and Mathematical Physics, under project {\it Quantum aspects of
gravity in cosmological models, phenomenology, and geometry of
space-time}. Many calculations where done by Symbolic Program REDUCE
3.8.}

\end{document}